\documentclass[twocolumn]{aastex631}
\usepackage{amsmath}
\usepackage{makecell}
\usepackage{CJKutf8}
\newcommand{\cntext}[1]{\begin{CJK}{UTF8}{gbsn}#1\end{CJK}\kern-1ex}
\graphicspath{{./}{figures/}}
\usepackage{booktabs}

\usepackage{soul}

\begin{document}

\title{Few Made It Out: A Multi-Messenger Study of an In Situ Solar Energetic Electron Event Driven by a Solar Jet}

\author[0000-0002-2633-3562]{Meiqi Wang}
\affiliation{Center for Solar-Terrestrial Research, New Jersey Institute of Technology, 323 Martin Luther King Boulevard, Newark, NJ 07102-1982, USA}

\author[0000-0002-0660-3350]{Bin Chen}
\affiliation{Center for Solar-Terrestrial Research, New Jersey Institute of Technology, 323 Martin Luther King Boulevard, Newark, NJ 07102-1982, USA}

\author[0009-0009-0282-4904]{Mallory Wickline}
\affiliation{Center for Solar-Terrestrial Research, New Jersey Institute of Technology, 323 Martin Luther King Boulevard, Newark, NJ 07102-1982, USA}

\author[0000-0003-2872-2614]{Sijie Yu}
\affiliation{Center for Solar-Terrestrial Research, New Jersey Institute of Technology, 323 Martin Luther King Boulevard, Newark, NJ 07102-1982, USA}

\author[0000-0002-2002-9180]{S\"{a}m Krucker}
\affiliation{University of Applied Sciences and Arts Northwestern Switzerland (FHNW), Bahnhofstrasse 6, 5210 Windisch, Switzerland}

\author[0000-0002-5865-7924]{Jeongwoo Lee}
\affiliation{Center for Solar-Terrestrial Research, New Jersey Institute of Technology, 323 Martin Luther King Boulevard, Newark, NJ 07102-1982, USA}

\author[0000-0002-5233-565X]{Haimin Wang}
\affiliation{Center for Solar-Terrestrial Research, New Jersey Institute of Technology, 323 Martin Luther King Boulevard, Newark, NJ 07102-1982, USA}

%% Note that the \and command from previous versions of AASTeX is now
%% depreciated in this version as it is no longer necessary. AASTeX 
%% automatically takes care of all commas and "and"s between authors names.

%% AASTeX 6.31 has the new \collaboration and \nocollaboration commands to
%% provide the collaboration status of a group of authors. These commands 
%% can be used either before or after the list of corresponding authors. The
%% argument for \collaboration is the collaboration identifier. Authors are
%% encouraged to surround collaboration identifiers with ()s. The 
%% \nocollaboration command takes no argument and exists to indicate that
%% the nearby authors are not part of surrounding collaborations.

%% Mark off the abstract in the ``abstract'' environment. 
\begin{abstract}

When \textit{in situ} solar energetic electron (SEE) events are closely associated with nonthermal flares, the escaping electron population is frequently observed to be much smaller than the nonthermal-radiation-emitting population near the solar surface. If a single accelerated population drives both signatures, the physical mechanism causing this severe deficit of upward-propagating electrons remains poorly understood.Focusing on one of the 2022 November 10--12 SEE events associated with recurrent solar jets and interplanetary type III radio bursts, we present a new, combined microwave--X-ray analysis using the Expanded Owens Valley Solar Array (EOVSA) and the Spectrometer/Telescope for Imaging X-rays (STIX) aboard Solar Orbiter. This synergy enables, for the first time for such an event, spatially resolved diagnostics over a broad energy spectrum of the near-Sun energetic electrons, complemented by \textit{in situ} measurements made by spacecraft at multiple heliocentric longitudes and distances. Consistent with earlier results based on \textit{in situ} and X-ray data, our results show that only 0.1--1\% of energetic electrons escape into interplanetary space. Crucially, the new microwave spectral imaging analysis suggests that energetic electrons are strongly concentrated in a compact region just above a mini-flare arcade at the base of the jet spire, and that their number density decreases by at least two orders of magnitude in the direction of the jet spire away from this region. This steep gradient, revealed by the microwave diagnostics, points to efficient local acceleration and trapping in the region analogous to the above-the-looptop ``magnetic bottle'' region in major eruptive flares, allowing only a small fraction of electrons to access open magnetic field lines and enter interplanetary space.

\end{abstract}

%The joint spectral analysis of microwave and hard x-ray emissions contributes to the broader spectrum of energetic electrons.

%% Keywords should appear after the \end{abstract} command. 
%% The AAS Journals now uses Unified Astronomy Thesaurus concepts:
%% https://astrothesaurus.org
%% You will be asked to selected these concepts during the submission process
%% but this old "keyword" functionality is maintained in case authors want
%% to include these concepts in their preprints.
\keywords{Solar flares(1496), Solar energetic particles(1491), Solar radio emission(1522), Solar x-ray emission(1536)}

%% From the front matter, we move on to the body of the paper.
%% Sections are demarcated by \section and \subsection, respectively.
%% Observe the use of the LaTeX \label
%% command after the \subsection to give a symbolic KEY to the
%% subsection for cross-referencing in a \ref command.
%% You can use LaTeX's \ref and \label commands to keep track of
%% cross-references to sections, equations, tables, and figures.
%% That way, if you change the order of any elements, LaTeX will
%% automatically renumber them.
%%
%% We recommend that authors also use the natbib \citep
%% and \citet commands to identify citations.  The citations are
%% tied to the reference list via symbolic KEYs. The KEY corresponds
%% to the KEY in the \bibitem in the reference list below. 

\section{Introduction}\label{sec:intro}

%general introduction on SEEs
Solar energetic electron (SEE) events involve the release of energetic electrons associated with solar activity, with energies ranging from a few keV to $\sim$MeV \citep{Reames1999}. A group of SEE events features relatively short duration, which are sometimes referred to as ``impulsive'' or ``prompt" SEE events. They are closely associated with magnetic reconnection processes occurring on the Sun and are typically accompanied by ${}^{3}\mathrm{He}$ enrichment \citep{Lin1996, Reames2021}. They have been reported to be closely associated with type III radio bursts, which are driven by energetic electrons traveling along open magnetic field lines \citep{Wang2012},
%discuss the ISEEs origin 
%The solar origin of $^3He-$rich impulsive solar energetic particle events remains an open question. 
and solar jets \citep{Wang2006, Nitta2006, Nitta2008, Krucker2011, Buvcik2020}. As both type III radio bursts and jets involve open field lines, it has been interpreted that energetic electrons can access and escape into interplanetary space through interchange magnetic reconnection with the ambient magnetic field lines \citep{Shibata1992, Shibata1996}. However, studies suggest that some impulsive SEE events are associated with coronal mass ejections (CMEs), extreme ultraviolet (EUV) waves, or other complex processes \citep{Wiedenbeck2013, Nitta2015, buvcik2015, Nitta2023}.

%This paragraph is about the electron spectra
%In addition to remote sensing observations near the solar surface, the spectra of in situ SEEs in the interplanetary medium also provide valuable information about their transport processes. \citet{Krucker2009} reported a statistical study on the spectra of SEEs observed by the WIND spacecraft in the energy range of $\sim$ 1 to 300 keV. They found that the electron peak flux spectra are generally show a broken power-law distribution with the break energy at $\sim$60 keV. The reason why causing the common observed broken power law remains unclear. \citet{Kontar2009} suggest, based on simulations, that low-energy electrons undergo significant energy loss due to wave-particle interactions. On the other hand, \citet{Strauss2020} report that pitch-angle scattering can influence electron spectral changes at higher energies ($>100$ keV). However, their observations are based on single-spacecraft measurements at 1 AU. The peak flux may vary significantly depending on the heliocentric distance \citep{Rodriguez-Garcia2023} and/or the magnetic longitude separation \citep{Pacheco2017}.

Joint remote-sensing observations and \textit{in situ} measurements provide a more comprehensive picture of the acceleration and transport of energetic electrons from near the solar surface to the interplanetary space. However, it remains an outstanding question whether the SEEs observed in the interplanetary space share the same origin as those emitting nonthermal microwave or HXR emissions near the solar surface. Multiple studies comparing the spectra of HXR-emitting energetic electrons near the surface with those from \textit{in situ} measurements have identified a positive correlation between their spectral indices, suggesting that they likely share a common origin \citep{Krucker2007, Wang2021, Dresing2021}. However, it is puzzling that these studies comparing the number of HXR-emitting and \textit{in situ} measured energetic electrons found that only a very small fraction of electrons, $\sim$0.1–1$\%$, escape into the interplanetary space, compared to the HXR-emitting population near the surface\citep{Krucker2007, Wang2021, Dresing2021, Wang2025}. This observed highly imbalanced partition significantly deviates from the scenario sometimes presumed, where particle acceleration occurs at a common site in the corona and ejects an equal number of energetic electrons in both upward and downward directions (see, e.g., the standard flare scenario depicted in \citealt{Shibata1995}).   
%reach the spacecraft at 1 AU. 
%However, the reason for such a small fraction of escaping electrons remains inconclusive.

%the idea of the previous study

One possible explanation is that initial acceleration occurs in the relatively high corona \citep{Wang2021}. Under this model, upward-propagating electrons escape directly into interplanetary space, while the downward-propagating electrons undergo an additional acceleration process in the newly closed field near the surface, leading to an increased number of X-ray-emitting electrons \citep{Krucker2007,Wang2021}. One of the main arguments favoring this multiple-acceleration-site scenario is that some of the observed SEEs have power-law spectra extending down to several keV, which are deemed inconsistent with pure injection from the low solar corona, since the low-energy electrons would have experienced significant Coulomb collisional loss \citep{Wang2021}.  Recent statistical studies \citep{wangw2023, wang2024} also argued that, despite a positive correlation, electron spectra derived from the thick-target HXR model are inconsistent with those of the \textit{in situ} electrons, which also appears to argue against the common acceleration site scenario. %Furthermore, \citet{Wang2021} employed a column collision model and suggested that the source of upward-propagating electrons is located high in the corona. 

The other scenario, recently suggested by \citet{Chen2024}, is that the accelerated electrons are strongly trapped in a nearly closed ``magnetic bottle'' region located below the reconnection X point, with only a small fraction managing to escape upward. This scenario naturally explains the highly imbalanced partition between the number of HXR/microwave-emitting electrons near the solar surface and the escaped \textit{in situ} electrons, as well as their positive correlation with each other. However, more detailed investigations are needed to account for other observed phenomena.

On the other hand, the estimation of the total number of escaping electrons can be significantly affected by energy-dependent transport effects \citep{Lin1985, Kontar2009, Strauss2020}. The electrons can experience Coulomb collisions during their propagation from the corona to the interplanetary space, with the lower-energy electrons enduring more significant losses \citep{Lin1985}. Furthermore, \citet{Kontar2009} propose that beam-driven plasma turbulence and solar wind density inhomogeneities play a critical role in producing the spectral break often observed in the electron spectra, resulting in flattening at lower energies (typically tens of keV). In addition to the various transport effects that influence low-energy electrons, \citet{Strauss2020} suggests that the energy loss of higher-energy electrons due to pitch-angle scattering contributes to spectral softening in the higher energy range. Thus, multiple spacecraft at various longitudinal and radial positions can provide valuable insight into the transport of energetic electrons in interplanetary space. They can also improve the estimation of the angular spread of the energetic electrons into interplanetary space and offer refinements on the estimations of the number of escaping electrons (see, e.g., \citealt{Xie2019}). %Alongside the comprehensive \textit{in situ} observations provided in this study, electron energy spectra from remote sensing over a broad energy range also enable effective comparison between remote-sensing and \textit{in situ} electron spectra.

%In addition to the joint microwave and HXR observations, \textit{in situ} measurements from multiple spacecraft at various longitudinal and radial positions provide valuable insights into the transport of energetic electrons in interplanetary space. Such multi-spacecraft observations enhance the estimation of the electron injection cone and offer further investigation of the factors contributing to the extremely small number of escaping electrons (see e.g,. \citealt{Xie2019}).

%what's new in this study about solving this problem
%Microwave imaging spectroscopy provides critical information about the possible electron acceleration sites and offers further insight into the total number of escaping electrons \citep{ Chen2013, Chen2018, wang2023}. \citet{Chen2018} utilized observations from the Karl G. Jansky Very Large Array (VLA) in the 1–2 GHz L band to derive detailed trajectories of type III-emitting electron beams, interpreting each beam-diverging site as a reconnection null point. In the meanwhile, the electrons can emit bremsstrahlung emission in the HXR range also provide the reconnection topology of the electron's origin from the sun in solar jets \citep{Krucker2011,Glesener2012}. \citet{Glesener2012} imaged the double coronal HXR source and suggested the electron acceleration in the interchange reconnection geometry. 

As for remote-sensing diagnostics of the energetic electrons near the solar surface, HXR bremsstrahlung emission provides excellent diagnostics of energetic electrons from $\sim$10--100 keV \citep{Brown1971}. However, due to the sensitivity limitation of current X-ray instrumentation, it remains challenging to routinely observe the X-ray spectra above $\sim$100 keV except for large events. In addition to bremsstrahlung radiation, non-thermal electrons gyrating in the coronal magnetic field can emit gyrosynchrotron radiation in the microwave range. Although microwave emission is contributed by a broad spectrum of energetic electrons, the peak contribution usually comes from the electron population at $\gtrsim$100 keV \citep{Krucker2020}. Thanks to their sensitivity to different energy regimes, joint analysis of both X-ray and microwave data is particularly powerful for constraining energetic electron distribution over a wide energy regime \citep{White2011,Chen2021}. Since 2017, the Expanded Owens Valley Solar Array (EOVSA; \citealt{Gary2018}) has provided microwave imaging spectroscopy spanning from 1 to 18 GHz with a 1-second time resolution, enabling spatially resolved spectral analysis near the acceleration site. The availability of microwave imaging spectroscopy, together with complementary X-ray observations, allows us to constrain the energetic electron population over a wide energy range, improving the comparisons to the \textit{in situ} measurements.

In this work, we investigate a series of SEE events associated with EUV solar jets observed by the Solar Orbiter (SolO; \citealt{Muller2020}) in mid-November 2022, a period of good magnetic connectivity between the spacecraft and the jets' source active region. These SEEs have been reported by \citet{Lario2024}, who conducted a detailed investigation into their relationship with EUV jets, type III radio bursts, HXR flares, and associated CMEs. They suggested that most of the SEE events exhibit a strong temporal and spatial association with HXR flares and EUV jets. %Additionally, they performed a detailed timing analysis to examine the possible time delays between the onset of SEEs, EUV jets, and HXR flares. 
Here, we extend the investigation of these SEE events by focusing on a subset of events associated with a solar active region that is magnetically connected to the SolO spacecraft. We then carry out a detailed case study on a specific SEE event associated with a blowout solar jet. By integrating HXR and microwave imaging spectroscopy from the Spectrometer/Telescope for Imaging X-rays (STIX; \citealt{Krucker2020}) onboard SolO and EOVSA, respectively, with multi-spacecraft \textit{in situ} measurements from various heliocentric longitudes and distances, we develop a more comprehensive understanding of SEE acceleration and transport.

Plan of this paper is as follows:
Section~\ref{SEE_events_overview} provides a brief overview of these SEEs in association with type III radio bursts and solar jets. Section~\ref{case_study_overview_sec} presents a focused study on the event that occurred on 2022 November 12. After first providing a brief introduction to the event and its context, we conduct a detailed analysis utilizing multi-wavelength remote-sensing data and \textit{in situ} measurements. Finally, in Section~\ref{discussion}, we summarize our main findings and discuss their implications, with a special focus on understanding why only a small fraction of energetic electrons can escape to the interplanetary space.

\section{The Context: A Series of SEE Events} \label{SEE_events_overview}
%\subsection{Overview} \label{overview}

\begin{figure*}
    \center
    \includegraphics[width=1.0\textwidth]{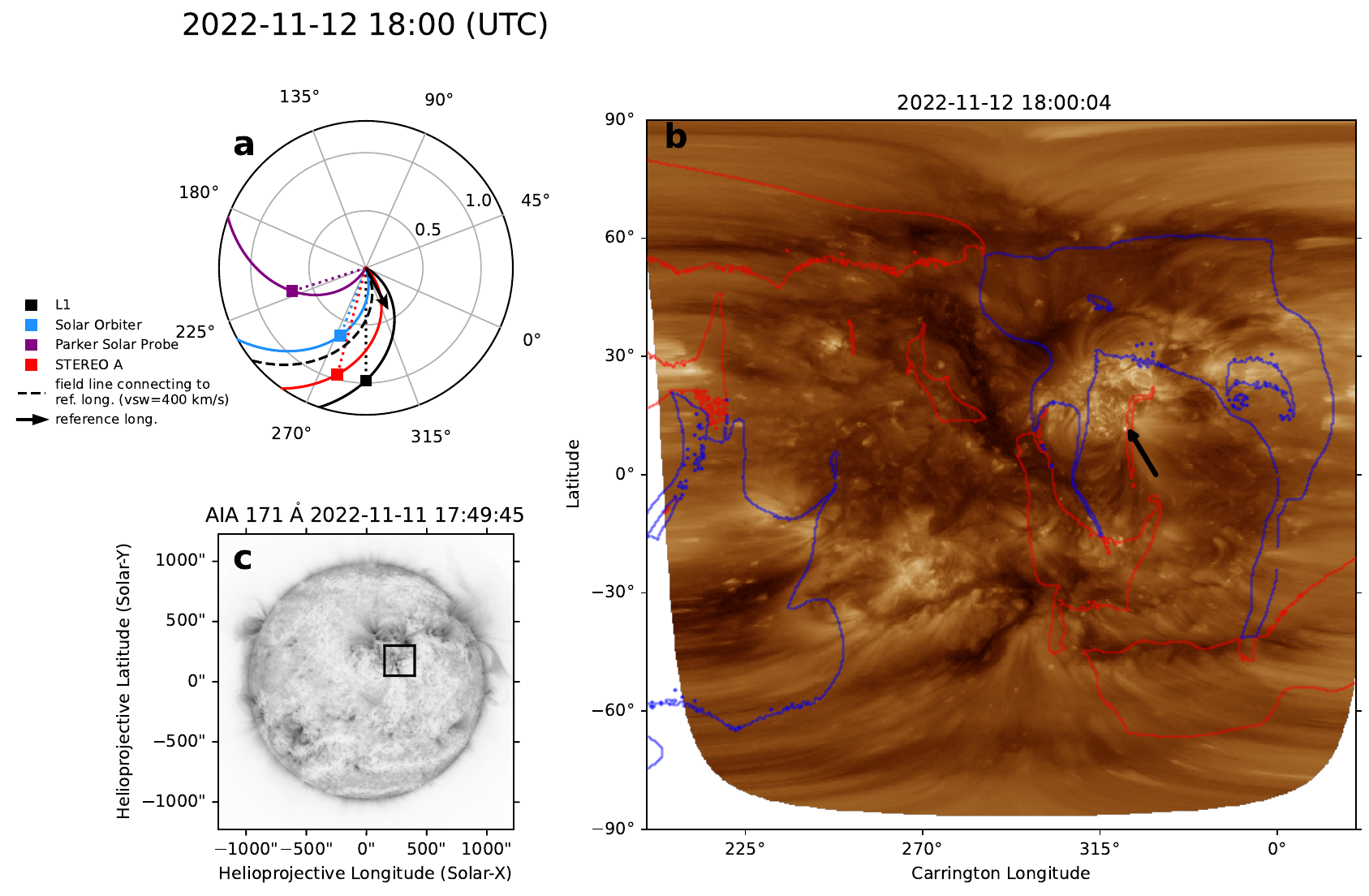}
    \caption{(a) Spacecraft configuration on 2022 November 12. The black arrow represents the longitude of AR 13141. The solar wind speed is assumed to be $400\ \mathrm{km}\ \mathrm{s}^{-2}$. (b) Open-closed field map of CR 2264 derived from the 3D MAS model. The background is SDO/AIA 193 \AA\ map reprojected into the Carrington coordinates. The red and blue contours mark the regions of open magnetic fields of positive and negative polarity, respectively. The black arrow indicates the jet region, which is associated with positive open magnetic field lines at the western edge of AR 13141. (c) Full-disk image from the SDO/AIA 171 \AA\ channel at 17:49:45 UT on 2022 November 11. The black rectangle encloses the AR 13141. } 
    \label{Fig_solar_mach_spacecraft}
\end{figure*}

\begin{figure*}
    \center
    \includegraphics[width=0.95\textwidth]{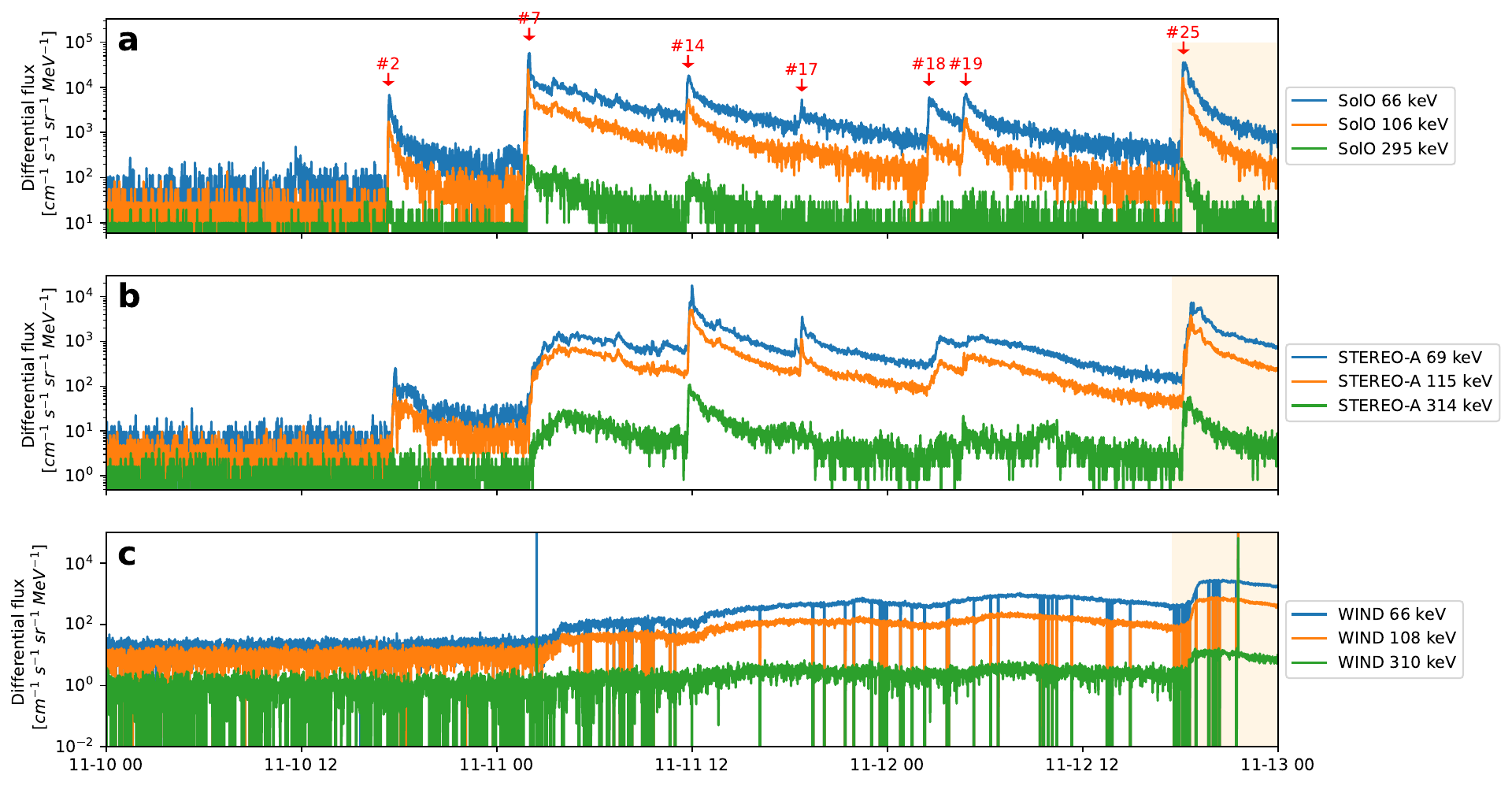}
    \caption{Electron differential flux observed by SolO/EPD (a), STEREO-A/SEPT (b), and WIND/3DP (c), respectively. The orange-shaded region shows the 2022 November 12 SEE event observed by both EOVSA and SolO/STIX, which is selected for spectral analysis. The seven red arrows with red numbers indicate the SEE events that can be clearly isolated in SolO/EPD's 66 keV channel, and are also marked as red in Figures~\ref{Fig_overview1}(a) and (e).}
    \label{in-situ_obs}
\end{figure*}

\begin{figure*}
    \center
    \includegraphics[width=1.0\textwidth]{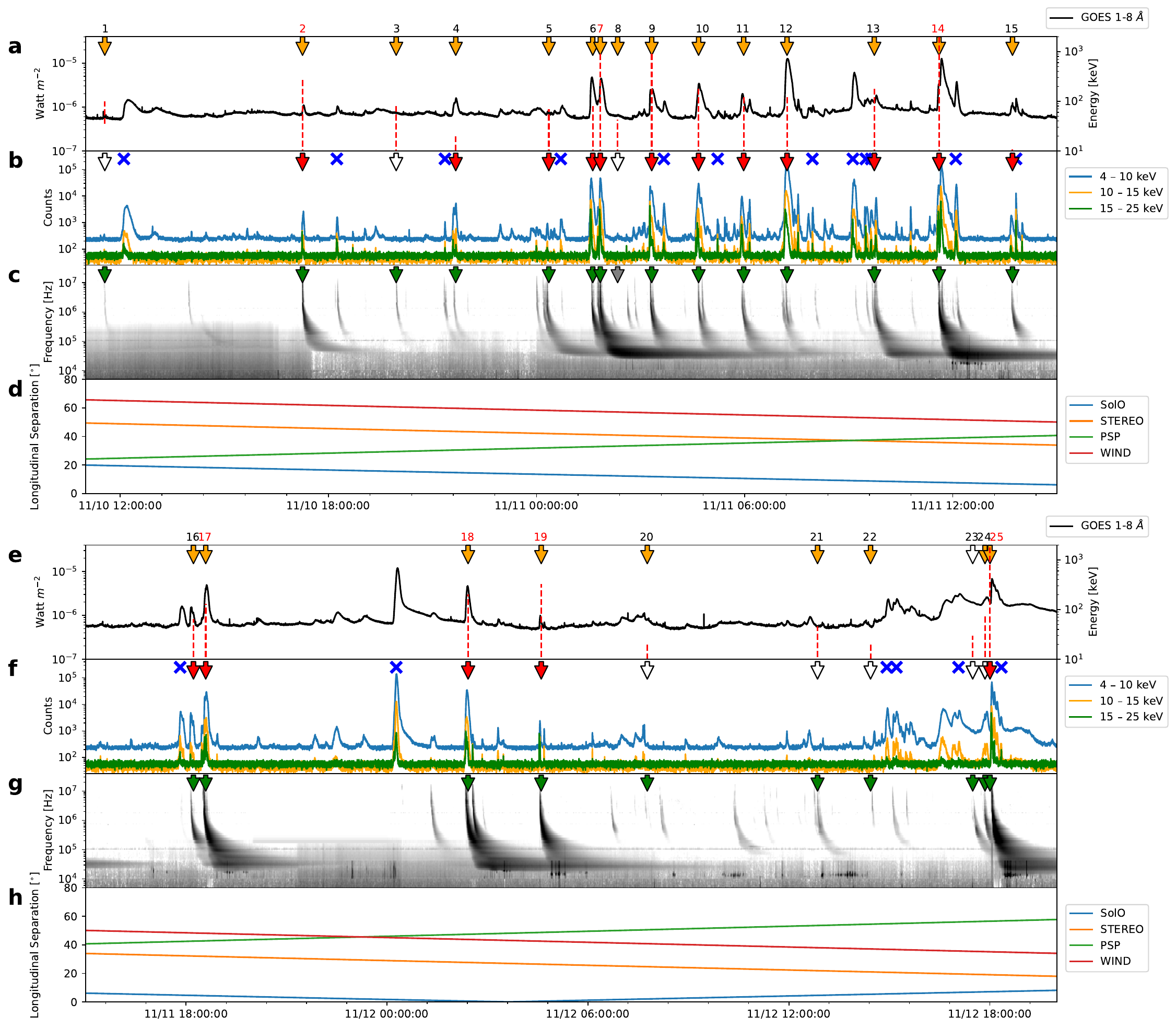}
    \caption{Relationship between the SEE events and GOES, STIX, and type III radio bursts during the period between 2022 November 10 at 11:00 UT and 2022 November 12 at 20:00 UT. The arrows labeled 1 to 25 indicate the times of the 25 SEE events identified by \citet{Lario2024}, among which seven of them marked as red can be clearly identified in the 66 keV channel observed by SolO shown as red arrows in Figure~\ref{in-situ_obs}. (a) GOES SXR light curve. The red dashed lines correspond to the SEEs event reported by \citet{Lario2024}, with the vertical heights representing the energy range of the SEEs. The arrows with (without)orange color mark the SEEs associated with (without) the EUV jets. (b) SolO/STIX light curve at three different energy channels. Red arrows indicate SEEs associated with X-ray flares that have corresponding STIX imaging data, while uncolored arrows mark SEEs without available STIX imaging data. Blue cross markers indicate STIX flares with imaging data that are not associated with SEEs. (c) Interplanetary type III radio bursts recorded by STEREO-A/WAVES. Green arrows indicate SEEs associated with type III radio bursts. All the SEE events but one (marked in a gray arrow), are associated with type III radio bursts. (d) Magnetic longitudinal separation between AR 13141 and SolO, PSP, WIND, and STEREO-A. (e)-(h) correspond to the same quantities as (a)-(d), but are presented over a different time interval.}
    \label{Fig_overview1}
\end{figure*}

%\begin{table*}[]
%\centering
%\begin{tabular}{llll}
%    Is the SEE event associated   &  & True (25) &  False %\\
%    \toprule
%    With an EUV flare?  &  &   25/122 (20.49\%)      & 97 (79.51\%)   \\
%    \toprule
%    With an X-ray flare?  & &   17/36 (47.22\%)      & 19 (52.78\%)   \\
%    \toprule
%    With a jet/eruption?  & True (99) &   24 (25.25\%)      & 73 (73.73\%)   \\
%    Without a jet/eruption & False (23) &    1 (4.34\%)     & 22 (95.65\%) \\
%    \toprule
%    With an IP type III burst?  & True (40) &   25 (62.50\%)      & 15 (37.50\%) \\
%     & False (82) &   0 (0.00\%)     & 82 (100.00\%) \\
%    \toprule
%    Both a jet/eruption \& an IP type III burst?  & True (37) &   24 (64.86\%)      & 13 (35.13\%) \\
%    \toprule
 %   Either a jet/eruption or an IP type III burst?  & ? & %  ?      & ? \\
% & False (20) &   0 (0.00\%)      & 20 (100.00\%) \\
%\toprule
%\end{tabular}
%\caption{The True-False Table regarding the relationship between the SEEs, EUV flares, STIX flares, Solar jets, and type III radio bursts. The number of events for each event type is indicated in parentheses.}
%\label{table1}
%\end{table*}

During the period from 2022 November 9 to 2022 November 14, the Energetic Particle Detector (EPD; \citealt{Rodriguez-Pacheco2020}) on board SolO observed a series of SEE events. Figure~\ref{Fig_solar_mach_spacecraft}(a) shows the locations of multiple spacecraft, including SolO, STEREO-A, and Parker Solar Probe (PSP) on 2022 November 12 (produced using the Solar Mach Python package; \citealt{Gieseler2022}). SolO was positioned at a distance of 0.61 AU from the Sun, while PSP was located at 0.67 AU. %Figure~\ref{Fig_solar_mach_spacecraft}(b) illustrates the results of a potential field source surface (PFSS) magnetic field model using the Carrington Rotation (CR) 2264 synoptic map obtained by the Helioseismic and Magnetic Imager (HMI; \textbf{add reference}) instrument onboard the Solar Dynamics Observatory (SDO) as the input, assuming a constant solar wind speed of 400~km~s$^{-1}$ and a source surface height at 2.5 solar radii. The PFSS magnetic model is derived using the Python package \textsc{pfsspy} \citep{Stansby2020}. The PFSS model results indicate that both PSP and SolO are magnetically connected to an active region group, which consists of two NOAA active regions AR 13140 and AR 13141. 

A total of 32 SEE events during this period were detected by SolO as previously reported by \citet{Lario2024}, who investigated the relationships among SEEs, solar jets, type III radio bursts, and STIX flares. % during the period from 9 to 14 November 2022. 
A detailed summary of the events is provided in Table~1 of \citet{Lario2024}. %In this study, we further explore SEEs and their associations with STIX flares, EUV brightenings, type III radio bursts, and jets. 
%A total of 307 events have been recorded across multiple active regions. 
%\textbf{This is the place to introduce the SEE events observed by SolO, STEREO, and WIND.}
%Figure~\ref{in-situ_obs}(a) shows the energetic electron flux obtained by SolO/EPT at three selected energy channels from 2022 November 10 to 2022 November 13. 
In addition to SolO/EPD, this series of SEE events was also recorded by the Wind 3D Plasma and Energetic Particle instrument  (3DP; \citealt{Lin1995}) on board WIND and the Solar Electron and Proton Telescope (SEPT; \citealt{Muller-Mellin2008}) on board STEREO-A. Figures~\ref{in-situ_obs}(a)–(c) display the electron flux in three selected energy channels as observed by SolO/EPD, STEREO-A/SEPT, and WIND/3DP, respectively. As described in Appendix B of \citet{Lario2024}, the sun aperture aboard STA/SEPT was not pointed toward the Sun during the event period. To facilitate a more accurate comparison of electron flux among different spacecraft, we employ sector-averaged flux for SolO/EPD and STEREO-A/SEPT, calculated as the mean of the electron flux measured by detectors oriented in different viewing directions. For WIND, we use the omnidirectional electron flux data that is already available. In the STEREO-A electron flux profiles shown in Figure~\ref{in-situ_obs}(b), multiple spikes are observed, each of which corresponds to an SEE event detected by SolO. 
%According to the magnetic longitudinal separation shown in Figure~\ref{Fig_overview}(d), WIND is poorly connected to the active region, with a separation exceeding $30^{\circ}$, which may explain the weak enhancements in the electron flux observed in Figure~\ref{in-situ_obs}(c). 

Out of the 32 SEE events, 25 are temporally associated with AR 13141 (Figure~\ref{Fig_solar_mach_spacecraft}(b)). This active region has been reported to be prolific in producing recurrent jets in its periphery region, possibly due to sunspot rotation \citep{Gou2024}. It also has a small longitudinal separation from the nominal footpoint of SolO (following the ballistic Parker spiral) during the period from 2022 November 10 to 12 (Figure~\ref{Fig_overview1}), suggesting a good magnetic connectivity to the SolO spacecraft. Recognizing the good magnetic connectivity to AR 13141, our study extends the work of \citet{Lario2024} by further focusing on these 25 SEEs and investigating their detailed spatial and temporal association with the recurrent jets, aiming to identify their specific physical origin. These events are also included in the CoSEE-Cat, an extensive catalog of SEEs observed by SolO \citep{Warmuth2025}.

To further investigate the magnetic field properties surrounding the recurrent jet region, we adopt a three-dimensional (3D) coronal model (Magnetohydrodynamic Algorithm outside a Sphere model; \citealt{Mikic1999}), which is derived using radial synoptic magnetograms computed from 720-s line-of-sight (LOS) magnetograms obtained by the Helioseismic and Magnetic Imager (HMI; \citealt{Scherrer2012}) onboard the Solar Dynamics Observatory (SDO) during Carrington Rotation 2264. The 3D model is then used to trace magnetic field lines from 1 to 30 $R_{\odot}$ from a dense grid of heliographic latitudes and longitudes, which are, in turn, used to evaluate areas with open or closed field lines. The resulting ``open-closed" field map\footnote{The open-closed map is synonymous with the ``coronal hole map'' terminology used on the Predictive Science website: \url{https://www.predsci.com/mhdweb/home.php}.} is shown in Figure~\ref{Fig_solar_mach_spacecraft}(b), with red and blue contours enclosing open field regions with positive and negative polarity, respectively. We find that the recurrent jet region is located in the close vicinity of a narrow open-field region, suggesting that SEEs associated with these jets may have access to the magnetic field opening to the interplanetary space. 
%We employed the open–closed field map at 18:00 UT on November 12, 2022, with the open field contours overlaid on the SDO/AIA 193 \AA image. The magnetic field lines were generated using the Magnetohydrodynamic Algorithm outside a Sphere (MAS) model \citep{Mikic1999}. This simulation is used a polytropic MHD model as input and solves the magnetohydrodynamic equations from 1 to 30 solar radii. The open positive magnetic field lines are located in the jet region on the western side of AR 13141 shown in Figure~\ref{Fig_solar_mach_spacecraft}(b). %\citet{Gou2024} reported that sunspot rotation may contribute to the triggering of solar jets, while \citet{Hou2023} reported a blowout jet event associated with a type II radio burst. 
%The event reported by \citet{Hou2023} is also included in the summary table presented by \citet{Lario2024} as being associated with a SEE event. However, this SEE event is unlikely to be related to the type II radio burst, as its onset was observed at SolO/EPD prior to the onset of the type II radio emission.

%All the 25 SEE events are associated with EUV brightenings located in a recurrent jet region in AR 13141 (outlined by the gray square in Figure~\ref{Fig_solar_mach_spacecraft}(b)). %The occurrence of multiple events associated with recurrent jets over the three-day period offers a favorable opportunity to investigate the source origins of these SEEs. 
%Considering further the excellent magnetic connectivity with SolO, this study will focus on these SEE events in AR 13141. 
Figure~\ref{Fig_overview1} presents the relationship between the 25 SEEs observed during the period from November 10 to 12 and the corresponding remote-sensing observations. SolO/STIX provides comprehensive coverage during this time. The STIX X-ray light curves in three different energy channels are shown in Figures~\ref{Fig_overview1}(b)/(f). The interplanetary type III radio bursts were recorded by multiple spacecraft, including Solar Orbiter/RPW and STEREO-A/WAVES, as shown in Figures~\ref{Fig_overview1}(c)/(g) (the SolO/RPW spectrum is not shown here). Figures~\ref{Fig_overview1}(d)/(h) show the magnetic longitudinal separation between the different spacecraft and AR 13141. Throughout the period, the separation between SolO and AR 13141 remains less than 20 degrees, reaching a minimum in the early hours of November 12. The magnetic footpoints of STEREO-A and WIND at 1 AU gradually approach the active region. The earliest clear enhancement observed by WIND, corresponding to event $\#7$ as marked in Figures~\ref{in-situ_obs}(a) and \ref{Fig_overview1}(a), occurred when WIND's magnetic footpoint had a longitudinal separation of $\sim57^{\circ}$ from the active region. As the separation becomes smaller and smaller at later times (Figure~\ref{Fig_overview1}(d) and (h)), this longitudinal separation may be regarded as an estimate for the maximum separation angle for WIND to detect this series of SEE events. During the same time period, PSP's magnetic footpoint is initially close to AR 13141 but moves progressively farther away over time. However, due to data gaps in the PSP observations during the time of interest, the PSP data are excluded from this study. The recorded SEEs across different energy ranges are indicated by vertical dashed lines in Figures~\ref{Fig_overview1} (a)/(e), most of which show good temporal correlation with the peaks of the GOES light curves.

%\subsection{The relationships between the EUV jets, STIX flares, type III radio bursts, and SEEs} \label{temporal_relation}

\begin{figure}
    \center
    \includegraphics[width=0.5\textwidth]{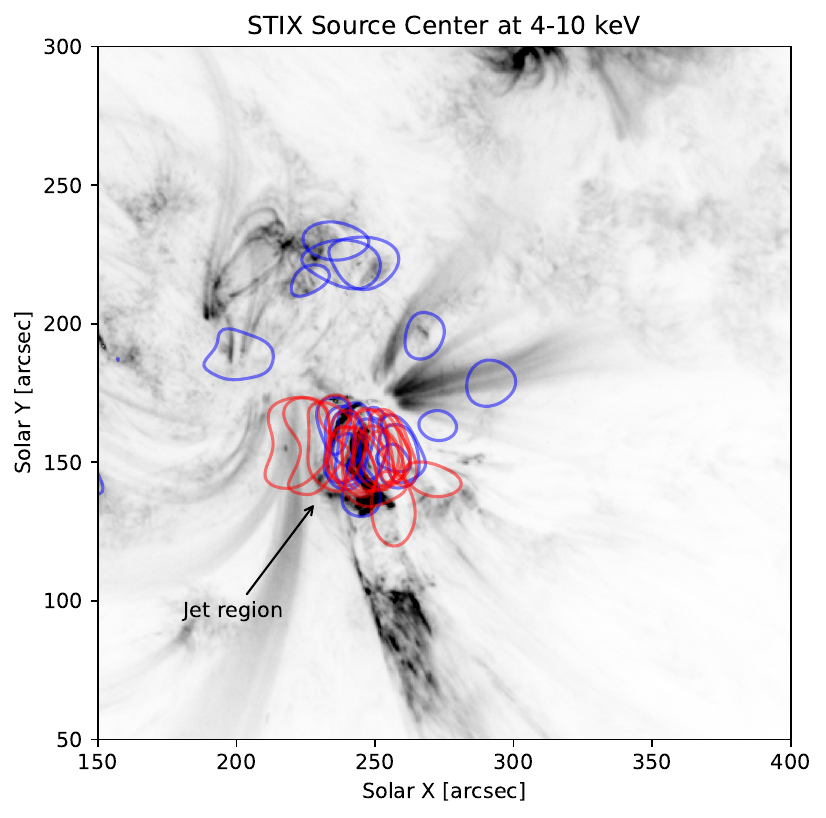}
    \caption{STIX sources at the 4--10 keV energy range (50$\%$ contour of the maximum intensity at each energy channel), overlaid with SDO/AIA 171 \AA\ image at 17:50 UT on 2022 November 11. The AIA field of view is the region corresponding to the black box shown in Figure~\ref{Fig_solar_mach_spacecraft}(b). The red contours correspond to STIX sources associated with SEEs, while the blue contours correspond to STIX sources without SEEs. }All STIX sources have been rotated to the SDO/AIA viewing perspective. 
    \label{Fig_flare_location}
\end{figure}

To better illustrate the relationships between these SEEs and various observational features, we mark the SEEs with different colors to highlight their associations with EUV jets (Figure~\ref{Fig_overview1}(a); yellow arrows), X-ray flares (Figure~\ref{Fig_overview1}(b); red arrows), and type III radio bursts (Figure~\ref{Fig_overview1}(c); green arrows). By closely examining the EUV images from Atmospheric Imaging Assembly (AIA; \citealt{Lemen2012}) on board the Solar Dynamics Observatory (SDO; \citealt{Pesnell2012}), we found that 24 out of the 25 SEEs are associated with jets, while only one event (marked as \#23) does not exhibit a clear jet signature.

%\citet{Lario2024} identified a total of 122 EUV events associated with AR13141, among which 36 have known STIX source locations with available pixel data. Here we have marked them in Figure~\ref{Fig_overview1} and Figure~\ref{Fig_overview1}(b). 
Figures~\ref{Fig_overview1}(b) and (f) shows the association of the SEE events with X-ray flares observed by SolO/STIX. Among the 25 SEE events, 8 events do not have STIX X-ray imaging data available\footnote{\url{https://datacenter.stix.i4ds.net/}}. The remaining 17 events with X-ray imaging data are indicated by red arrows in Figures~\ref{Fig_overview1}(b) and (f). The source locations of these X-ray flares are shown in Figure~\ref{Fig_flare_location}. The source images are co-aligned with the SDO/AIA 171 \AA\ image at a fixed reference time (17:50 UT on November 11), accounting for solar rotation. The red contours indicate STIX sources associated with SEEs, all of which are exclusively clustered around the recurrent jet region, as shown in Figure~\ref{Fig_overview1}. In contrast, STIX flares without associated SEEs are distributed not only within the jet region but also in other areas surrounding it. The source locations of STIX flares associated with both EUV jets and SEEs are tightly concentrated within the jet region, suggesting that these SEEs are likely to originate from the region with the recurrent jets.

Furthermore, nearly all SEEs are associated with interplanetary type III radio bursts, except that one event was detected with a delay of more than ten minutes relative to the corresponding type III burst (marked in gray in Figure~\ref{Fig_overview1}(c)). We further examine the events reported in Table 1 of \citet{Lario2024}, specifically those in AR 13141 but without jets/eruptions nor type III radio bursts. For these events, no SEEs are detected. The close association of the SEE events with jets and type III radio bursts in this active region suggests that, with good magnetic connectivity to the spacecraft and access to open field lines, the presence of solar jets and type III radio bursts can be effectively considered a necessary condition for the detection of SEEs.

\section{The 2022 November 12 Event} \label{case_study_overview_sec}

In Section~\ref{SEE_events_overview}, we have established that the recurrent jets at the edge of AR 13141 have excellent spatial and temporal correlation with SEE events observed by SolO. They also have direct magnetic connectivity to the spacecraft and access to open field lines. To further elucidate the relation between the HXR/microwave-emitting energetic electrons near the solar surface and those escaped to the interplanetary space, in this section, we will focus on a particular SEE-associated jet event that occurred on 2022 November 12 around 18:03 UT, with comprehensive coverage by multiple instruments providing multi-wavelength remote-sensing observations in microwave, (E)UV, and X-rays, together with \textit{in situ} measurements made by multiple spacecraft including SolO, WIND, and STEREO-A.

\subsection{Jet Driven by an Erupting Mini-Filament}

\begin{figure*}
    \center
    \includegraphics[width=0.95\textwidth]{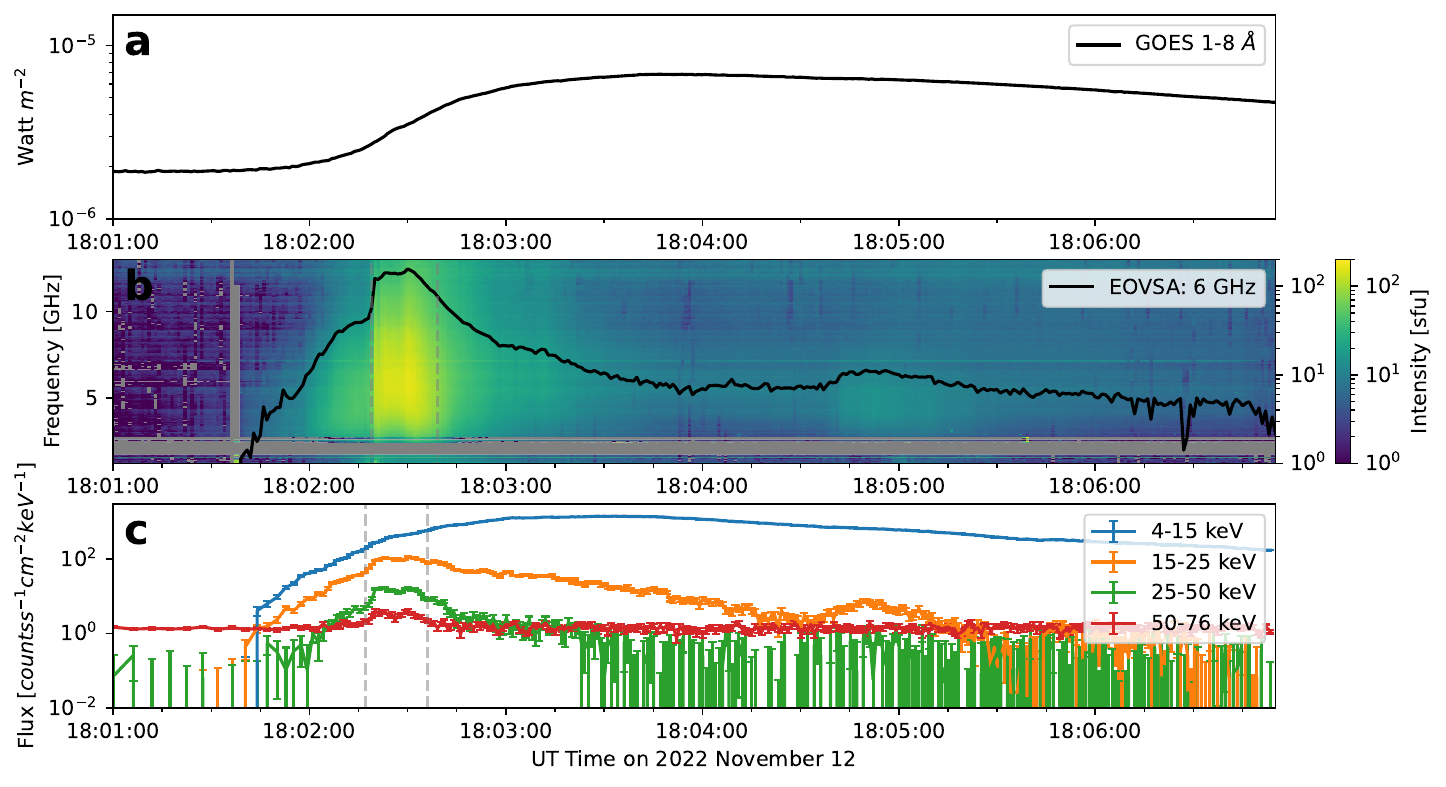}
    \caption{Overview of the time history of the 2022 November 12 jet event: (a) GOES SXR light curve. (b) Total-power radio dynamic spectrum in the microwave range observed by EOVSA, with the black curve representing the light curve at 6 GHz (the data gap region is indicated by a gray background). The gray dashed lines indicate the full-width-at-half-maximum (FWHM) duration of the 6 GHz microwave emission, which lasts 21 s. (c) STIX-observed HXR light curves at different energy channels. The pre-flare background has been subtracted. The gray dashed line indicates the FWHM duration of the 25–50 keV burst, which lasts 19 s.}
    \label{case_study_overview}
\end{figure*}

\begin{figure*}
    \center
    \includegraphics[width=0.95\textwidth]{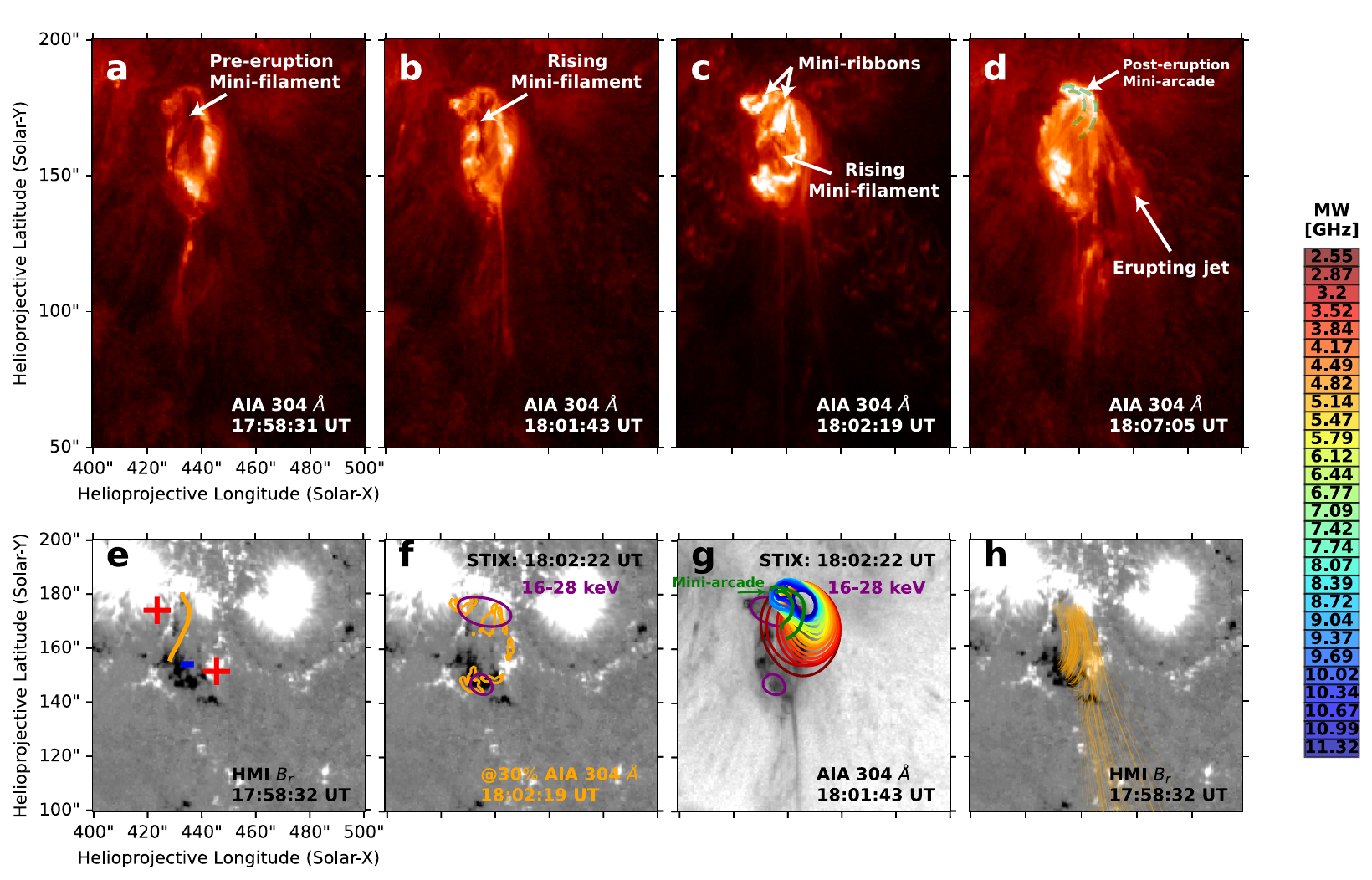}
    \caption{Morphology and evolution of the 2022 November 12 jet event. (a)–(d): Time-series EUV images observed by the SDO/AIA 304 \AA\ band. (e): SDO/HMI radial field magnetogram at 17:58:32 UT. The orange curve represents the mini-filament feature identified in panel (a). (f): The background shows SDO/HMI radial field magnetogram at 17:58:32 UT. The orange contours represent 30\% of the peak value from the SDO/AIA 304 \AA\ band at 18:02:19 UT in (c). The STIX nonthermal (16–28 keV) sources, shown in purple with 70\% contours, %and thermal (4–10 keV) sources, shown as red contours at the 50\% and 70\% levels, 
    have been reprojected onto the SDO/AIA viewing perspective. (g): Microwave sources with $50\%$ contours at different frequency channels, represented by different colors, overlaid on the SDO/AIA 304 \AA\ image. The purple contours represent the same STIX 16-28 keV sources as in panel (f). The green curve, derived from the green dashed lines in panel (d), shows the mini-arcade structure observed in AIA. (h): Selected field lines (yellow curves) derived from a potential magnetic field extrapolation model overlaid on the SDO/HMI radial field magnetogram.}
    \label{case_study_geo}
\end{figure*}

\begin{figure*}
    \center
    \includegraphics[width=0.95\textwidth]{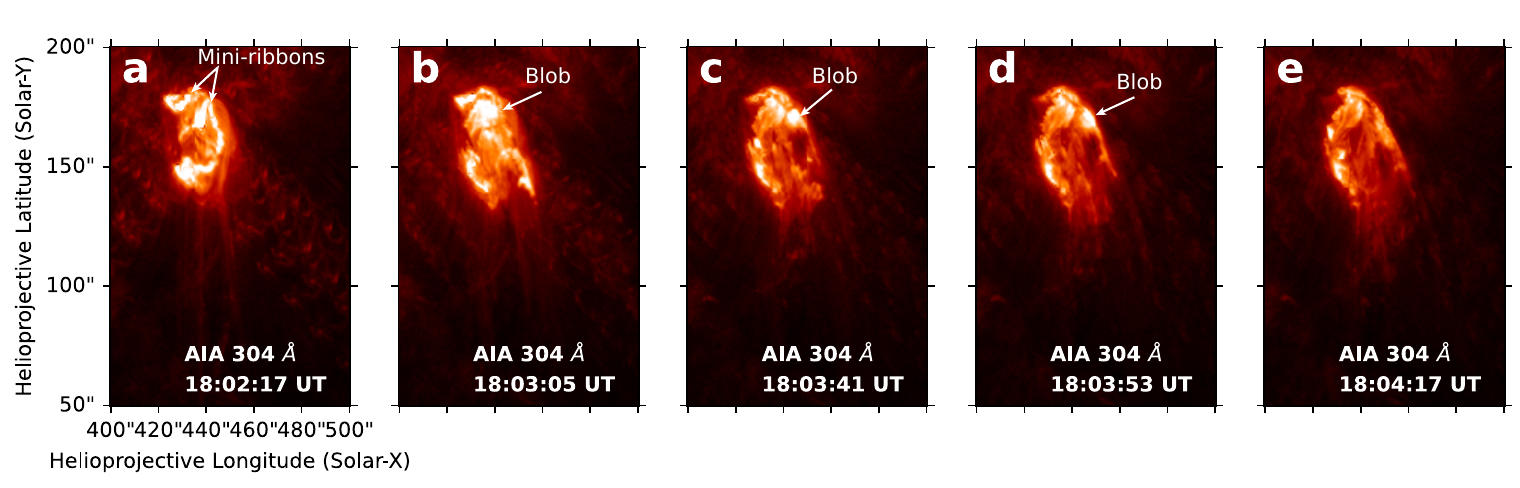}
    \caption{(a)–(e): Temporal evolution of the 2022 November 12 jet event featuring an erupting blob along the jet spire. The two white arrows in (a) indicate the two mini-ribbons. The white arrows in (b)--(d) indicate the erupting blob. }
    \label{reconnection_signature}
\end{figure*}

%In Section~\ref{in_situ_spectra_mul}, we discussed a SEE event observed by multiple spacecraft in the interplanetary space on November 12, 2022. This SEE event was well captured by SDO in EUV wavelengths, EOVSA in microwaves, and by SolO/STIX in HXRs. In this section, we will discuss results from a detailed analysis based on remote sensing observations associated with this SEE event. 

The solar counterpart of this SEE event is classified as a C6.6-class flare accompanied by an EUV jet observed by the Geostationary Operational Environmental Satellite (GOES; \citealt{Garcia1994}) spacecraft. Figures~\ref{case_study_overview}(a)-(c) show the soft X-ray (SXR), microwave, and HXR light curves from GOES, EOVSA, and STIX, respectively. %This event is classified as a GOES C6.6-class flare, accompanied by an EUV jet. 
%The HXR light curve above 25 keV, in the nonthermal energy range, exhibits a similar enhancement to the microwave emission during the peak.
Figures~\ref{case_study_geo}(a)-(d) show the jet eruption process observed by multiple channels of SDO/AIA. Just before the eruption, a mini-filament, appearing as dark material, marked by the white arrow, is visible in the SDO/AIA 304 \AA\ channel. About two minutes later, the filament begins to rise and erupt outward toward the southwest direction, shown in Figure~\ref{case_study_geo}(b). At the peak of the HXR and microwave light curves around 18:02:19 UT, two mini-ribbons appear on the northern side, indicated by two white arrows in Figure~\ref{case_study_geo}(c). The 30\% contours of the 304 \AA\ image, representing the ribbons, are overplotted on the radial magnetic field map in Figure~\ref{case_study_geo}(f). At about 18:04:42 UT, plasma materials have been ejected, forming the jet spire from the northeast to the southwest, as shown in panel (d). The pre-eruption dark filament marked by the yellow solid line is identified from the dark feature located between regions of opposite magnetic polarity, as shown in panel (a). The observed morphology in our case, featuring an erupting mini-filament from the base region, is consistent with the ``blowout-jet'' scenario \citep{Moore2010, Sterling2015}. The yellow lines in Figure~\ref{case_study_geo}(h) represent magnetic field lines derived from a potential force-free field extrapolation model (using the codes available in the \texttt{GX\_Simulator} package under \texttt{SSWIDL} \citep{Stupishin2020, Nita2023}. The field lines toward the south correspond to the geometry of the jet spire shown in Figure~\ref{case_study_geo}(d).

The temporal evolution of the event during the flare, as observed by the SDO/AIA 304 \AA\ channel, is shown in Figure~\ref{reconnection_signature}. Around the HXR and microwave peak, a bright blob appears (marked by the white arrow in panels (b)–(d)) and quickly erupts along the direction of the jet spire. %remains as a cusp-shaped structure located on the western side of the mini-ribbons. 
Such a small-scale eruption likely represents a portion of the destabilized filament near its northern footpoint, which erupts first to induce the two conjugate mini-ribbons in a fashion similar to the standard two-ribbon flare picture. In this case, magnetic reconnection may occur at a location in a region underneath the eruption, driving the energy release and particle acceleration during the microwave/HXR peak of the event.     %This brightening feature may represent a possible reconnection signature between the flux rope and the surrounding field lines. The potential reconnection region is indicated by a yellow dashed circle.

Due to SolO's different viewing perspective from Earth, we reproject STIX to match the viewing perspective of the AIA/HMI images. Figure~\ref{case_study_geo}(f) and (g) show the STIX sources during the event peak in the 16--28 keV (purple) energy channels, respectively. The viewing perspective of SolO/STIX results in a large viewing angle relative to the flare site. The foreshortening effect may lead to compromised resolution in the east-west direction for STIX imaging. Nevertheless, the northern 16–28 keV HXR source appears elongated between the positive and negative magnetic polarities and is aligned with the northern mini-arcade, shown in green. %The lower-energy 4–10 keV emission exhibits two sources, with the northern source also appearing to be elongated. 

EOVSA observed the full event in the 1–18 GHz microwave range at a 1-s time cadence \citep{Gary2018}. We perform imaging spectroscopy in the frequency range from 2.55 GHz to 11.32 GHz, which exhibits a strong signal-to-noise ratio. The EOVSA images have been calibrated, self-calibrated, and reconstructed using the \texttt{tclean} task available in the \textsc{CASA} package \citep{McMullin2007}. The frequency-dependent circular restoring beam has a size of 80\arcsec/$\nu_{GHz}$, where $\nu_{GHz}$ is the frequency in GHz. Figure~\ref{case_study_geo}(g) shows that the EOVSA multi-wavelength microwave sources at the peak time ($\sim$18:02:22 UT), with blue to red color contours representing decreasing frequencies. The microwave sources appear to be located close to the northern HXR source. %and should be located at above-the-loop-top region, as seen in panel (f), 
Particularly, the higher-frequency sources (cyan to blue colors) display an arcade-like shape bridging the conjugate mini-ribbons, which are located next to (or just above) the mini-arcade seen in EUV.  

%\subsection{Electron Spectra from Remote Sensing Observations: Joint Spectral Analysis of HXR and MW} %\label{HXR_mw_spectra}

\begin{figure*}
    \center
    \includegraphics[width=1.0\textwidth]{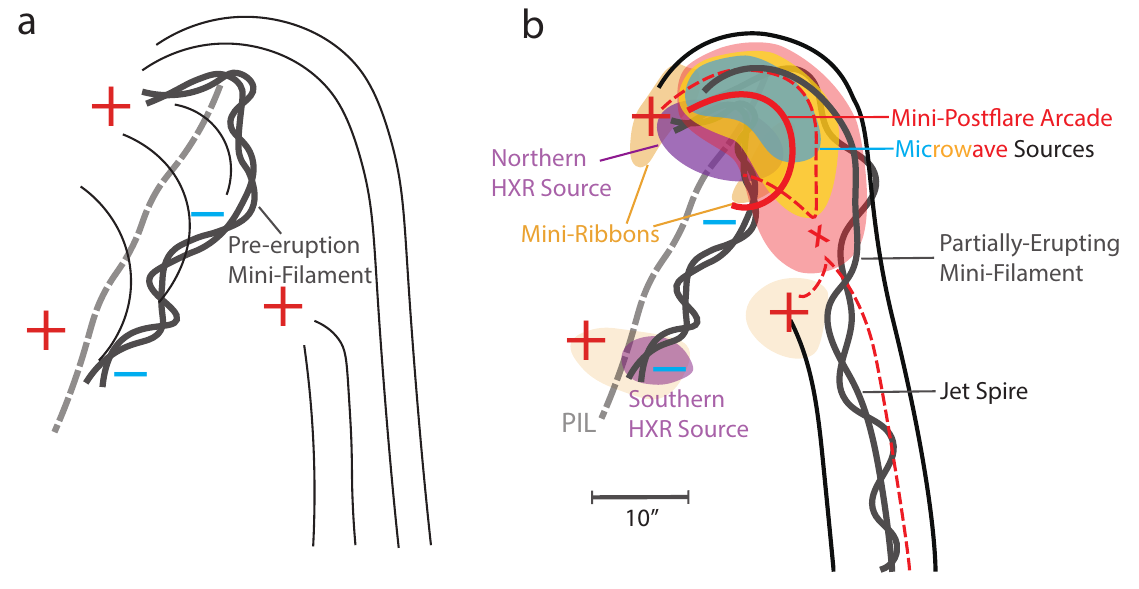}
    \caption{An illustration of the physical scenario of the blowout jet event. (a): Initially, a mini-filament lies close to the polarity inversion line. The black curves illustrate the pre-eruption magnetic field lines. (b) During the event, magnetic reconnection occurs underneath the partially-erupting filament material, leading to energy release and particle acceleration. A pair of conjugate mini-ribbons appears at the footpoint of the mini-postflare arcade, coinciding with an elongated HXR source that may be underresolved due to the foreshortening effect. High-frequency microwave sources (depicted in blue-shaded region), which display an arcade-like shape, are located above the mini-postflare arcade, while the low-frequency microwave sources (depicted in yellow-pink color) extend upward and display a cusp shape toward the presumed magnetic reconnection site. The location of the microwave sources are analogous to those observed in the above-the-looptop region in major flares. %the The HXR sources are elongated between the positive and negative polarities. 
    %The flare arcade on the northern side is visible in the SDO/AIA images.  
    The remote HXR footpoint may correspond to the southern footpoint of the mini-filament. Meanwhile, jet material flows along the field lines toward the southwest. %The brown dashed lines show the slit for further MW spectra analysis. 
    }
    \label{fig:cartoon}
\end{figure*}

Here, we summarize the multi-wavelength observations and illustrate our understanding of the event geometry in Figure~\ref{fig:cartoon}. Before the event, the mini-filament lies near and slightly above the polarity inversion line (PIL), marked by the gray dashed lines. %The blue lines represent selected magnetic field lines. 
During the initiation of the event, the mini-filament is activated, with its northern end undergoing a more rapid rise. Similar to the standard two-ribbon flare scenario, the rising filament induces magnetic reconnection underneath it, leading to the formation of a mini-postflare arcade bridging the two conjugate mini-ribbons seen in EUV (c.f., Figures~\ref{case_study_geo}(c) and (d)). Despite a compromised angular resolution due to foreshortening, the location of the two EUV mini-ribbons agrees with the elongated northern HXR source, suggesting the presence of precipitated nonthermal electrons as a result of the reconnection. The location of the southern HXR source agrees with the southern end of the pre-eruption mini-filament (c.f., Figure~\ref{case_study_geo}(f)), which may be due to the transport of nonthermal electrons along the magnetic flux rope toward its footpoints (e.g., \citealt{Chen2020fluxrope,Stiefel2023}). Meanwhile, the high-frequency microwave sources, which display an arcade-like shape, are located close to and seemingly just above the EUV mini-arcade (cyan and blue contours in Figure~\ref{case_study_geo}(g)). The low-frequency microwave source, on the other hand, extends beyond and above the mini-arcade toward the presumed reconnection site. This observed signature is analogous to previous observations of major eruptive flares in which microwave sources are present in the above-the-looptop region \citep{Chen2020,Yu2020}. They have been suggested to be the signature of nonthermal electrons accelerated and trapped in the ``magnetic bottle'' structure in this region \citep{Chen2024}. %During the peak time, when the microwave and HXR emissions reach their maxima, magnetic reconnection occurs between regions of opposite polarity. 
%The brightening cusp-shaped structure appears between the mini-arcade and the southern positive polarity, and may represent a possible magnetic reconnection signature between the flux rope and the surrounding ambient field lines. 
%A mini-scale flare arcade subsequently forms on the northern side as a result of this reconnection and is evident in the AIA channels, as indicated by a white arrow in Figure~\ref{case_study_geo}(c). 
%Two nonthermal HXR sources are located at the northern and southern footpoints. The northern HXR source extends across regions of positive and negative magnetic polarity. 
%At the peak time, the microwave sources, indicated by the blue-shaded area, show higher-frequency components extending toward the northern leg of the flare arcade. 
%The double structure of the higher-frequency microwave sources and the elongated HXR footpoint source supports the mini-arcade structure. 
The energy release induced by the reconnection drives a mixture of heated coronal plasma and cool filament material flowing along the open field lines, forming the observed jet spire toward the southwest. %The open-closed map in Figure~\ref{Fig_solar_mach_spacecraft}(b) indicates that the open field lines are located on the western side of the jet region. 
Complex internal and external magnetic reconnection can also occur during the filament eruption, enabling energetic electrons to escape into interplanetary space along the open field lines.

\subsection{Microwave and HXR Spectral Analysis} \label{spectra analysis}

\begin{figure*}
    \center
    \includegraphics[width=0.95\textwidth]{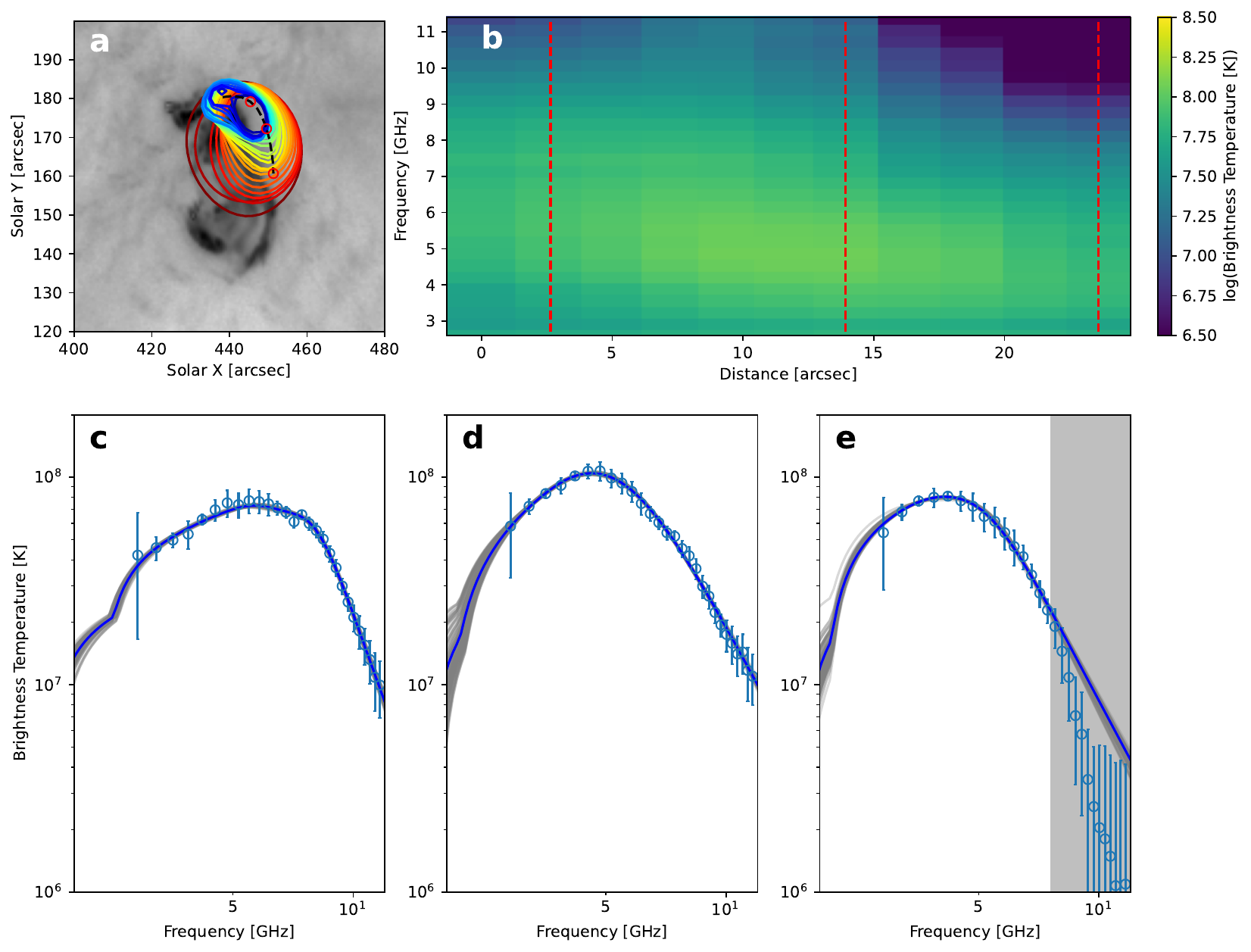}
    \caption{Microwave imaging spectroscopy during the peak of the eruptive jet event at 18:02:22 UT. (a): Microwave sources at different frequency channels (50\% contour of the respective maximum) are shown. The black dashed line represents the selected slit to generate the ``frequency-distance'' spectrogram in (b), %The red points mark the representative points indicated by the red dashed lines in (b). (b): The spectrum of the selected slit. 
    in which the $x$-axis shows the distance from the northernmost starting point, and the $y$-axis shows the frequency. The red dashed lines in (b), which correspond to the red circles in (a), indicate the three selected representative locations from which the microwave spectra shown in (c)-(e) are derived.%: The three representative microwave spectra located at three different positions along the slit, marked by the red dashed lines in Figure~\ref{MW_spectra}(c) and red points in (a)-(d). 
    The error bars in panels (c)–(e) represent three times the root mean square (RMS) of the source-free background. The gray curves show the MCMC runs, while the blue curves represent the best-fit results. The gray areas represent the frequency channels with insufficient signal-to-noise values and are therefore excluded from the spectral fitting. }
    \label{MW_spectra}
\end{figure*}

\begin{figure*}
    \center
    \includegraphics[width=0.9\textwidth]{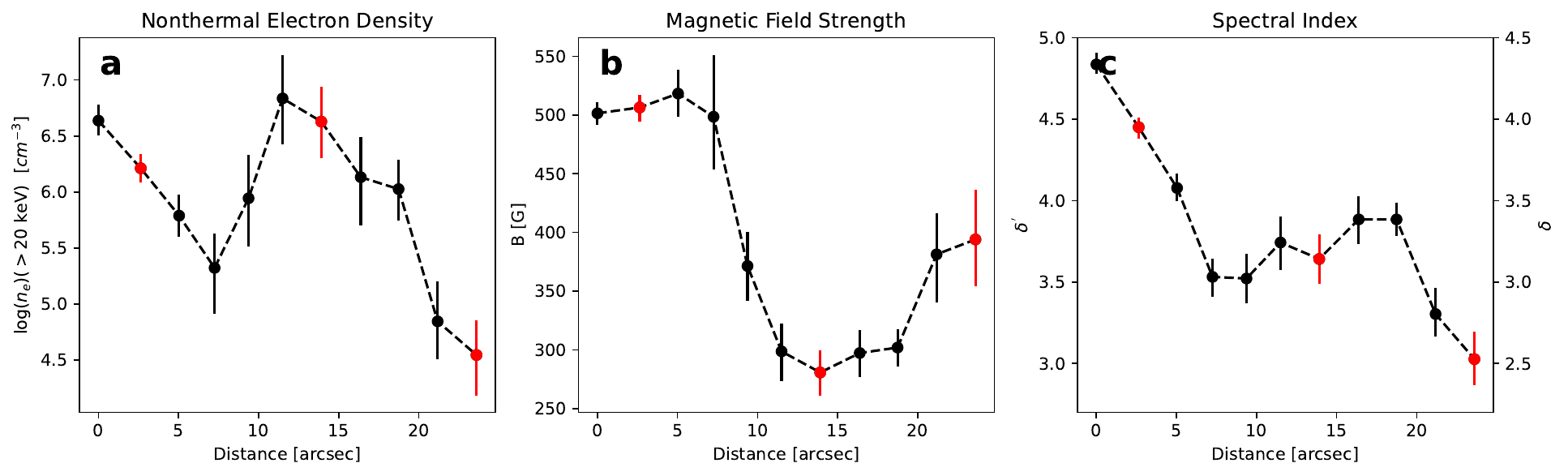}
    \caption{Best fit values for total nonthermal electron density $n_e$ ($>$20 keV), magnetic field strength $B$, and power-law index of the nonthermal electron density distribution $\delta'$ from the spatially resolved microwave spectral analysis. The error bars shown in the plots are estimated using the 1-$\sigma$ variance from the MCMC runs. %The three fitted parameters with 1-sigma error bars along the slit. 
    The three red points in each panel indicate the representative points for spectral analysis as in Figure~\ref{MW_spectra}. }
    \label{MW_spectra_results}
\end{figure*}

%\begin{figure*}
%    \center
%    \includegraphics[width=0.85\textwidth]{SEE_1800_event.pdf}
%    \caption{(a): The electron spectra from the selected event. The circle, triangle, and square markers with error bars indicate the electron spectra observed by SolO, STEREO-A, and WIND, respectively. The length of the error bars represents the 3-sigma uncertainty of the event background. The solid lines represent the spectral fitting results using a double power-law distribution. (b): The peak flux of the event observed by SolO (blue), STEREO-A (green), and WIND (red) at 120 keV. The peak flux has been corrected by $R^2$. The solid line represents the fitted Gaussian model, while the gray dashed lines indicate the fitted Gaussian model under the assumption that the peak flux is corrected by $R^3$.}
%    \label{Fig_instu_electron}
%\end{figure*}

\begin{figure}
    \includegraphics[width=0.5\textwidth]{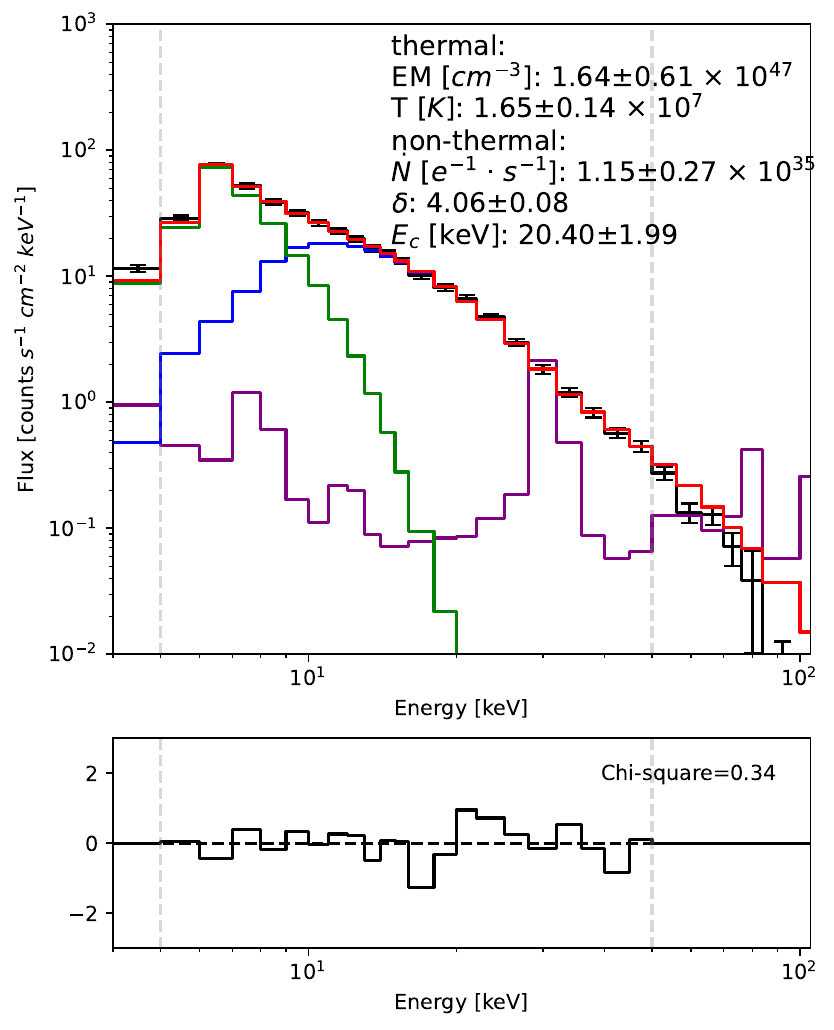}
    \caption{X-ray spectra fitting results based on SolO/STIX observations at 18:02:22 UT integrated over ten seconds. The black curve shows the data. The blue, green, and red curves show the best-fit nonthermal (assuming thick-target model) component, isothermal component, and the two components combined, respectively. The purple curve represents the background. %the yellow curve shows the fitted nonthermal component assuming the thick target model, and the green curve shows the fitted thermal component. 
    %The red curve represents the combination of the thermal and nonthermal fittings. 
    %The selected fitting time range is averaged over 5 seconds around the HXR peak in the 25–50 keV energy band. 
    The energy range selected for fitting is between 5 keV and 50 keV, as indicated by the gray dashed lines. The best-fit parameters are listed in the top-right corner of the plot (EM: emission measure. $T$: plasma temperature. $\dot N$: total nonthermal electron rate. $\delta$: spectral index of the nonthermal electron flux distribution.  $E_c$: low energy cutoff of the nonthermal electron distribution. }
    \label{hxr_spectral_fitting}
\end{figure}

\begin{figure*}
    \center
    \includegraphics[width=0.85\textwidth]{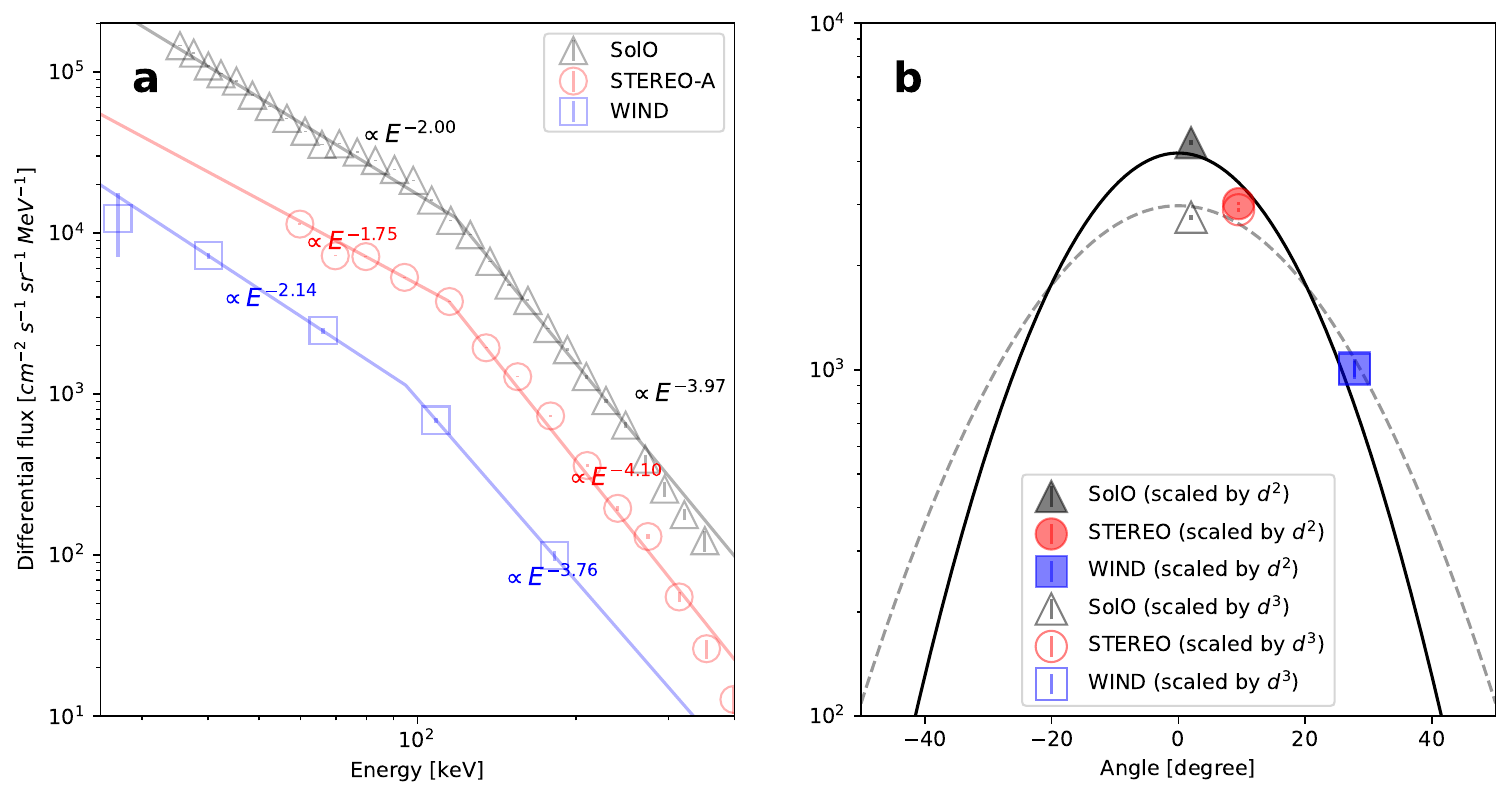}
    \caption{(a): \textit{In situ} energetic electron spectra from the 2022 November 12 SEE event. The circle, triangle, and square markers with error bars indicate the peak energetic electron spectra observed by SolO, STEREO-A, and WIND, respectively. The length of the error bars represents the 3-sigma uncertainty of the event background (most are too small to see). The solid lines represent the spectral fitting results using a double power-law model. (b): Peak 120 keV SEE flux as a function of magnetic longitudinal separation from the jet region, of the event observed by SolO (black), STEREO-A (red), and WIND (blue). Filled and open symbols show the results scaled using the $F\propto d^{-2}$ and $F\propto d^{-3}$ relation, respectively. Solid and dashed curves show the respective Gaussian fit results.}
    \label{Fig_instu_electron}
\end{figure*}
 
%Here we adopt a remote sensing approach, utilizing hard X-ray and microwave emissions from a specific SEE event to explore the nonthermal electron spectra at the acceleration site before their release. In Section~\ref{case_study:geo}, we investigate the jet eruption in the EUV wavelengths and suggest that the event corresponds to a blowout jet associated with a mini-filament eruption. Here, we illustrate this scenario in Figure~\ref{MW_spectra}(a). 

Microwave data recorded by EOVSA in this event enable us to perform spatially resolved spectral analysis at different locations. Here we select microwave spectra derived from one slit, starting from near the northern footpoint and extending southwest, as indicated by the black dashed lines in Figures~\ref{MW_spectra}(a). To ensure sufficient signal-to-noise for spectra analysis, we choose frequency channels where the peak flux is at least three times higher than the root mean square of a background, source-free region. The spatially-resolved brightness temperature spectra derived from the selected slit is shown in Figure~\ref{MW_spectra}(b), with the horizontal axis indicating the distance along the starting point of the slit (near the northern footpoint) and the vertical axis representing the microwave frequency. Three representative spectra, taken at distances of 2.7\arcsec, 14.0\arcsec, and 23.6\arcsec\ from the northern starting point (marked by the red dashed lines in Figure~\ref{MW_spectra}(b) and the red circles in Figure~\ref{MW_spectra}(a)), are shown in Figures~\ref{MW_spectra}(c)--(e). %correspond to three different locations: northern loop leg, above-the-looptop region, and the presumed reconnection X point. 

%The corresponding spectra are shown in Figures~\ref{MW_spectra_results}(e)–(g). 
The observed microwave brightness temperature spectra exhibits a positive slope at lower frequencies. After reaching the turnover frequency, it transitions to a negative slope at higher frequencies. Such a spectral shape is consistent with gyrosynchrotron radiation \citep{Dulk1985}, in which the low-frequency, positive--slope portion of the spectra is dominated by emission in the optically thick regime and the high-frequency, negative-slope portion is due to optically thin emission. Here, we adopt the fast gyrosynchrotron codes developed by \citet{Fleishman2010} and fit the spectra under the assumption of a homogeneous source with a power-law nonthermal electron distribution. The background plasma temperature is fixed to 3 MK. The low-energy cutoff of the nonthermal electron distribution is set to 20 keV, and the high-energy cutoff is set to 10 MeV\footnote{We note the spectral fit results are insensitive to the selected value of the high-energy cutoff, as long as it is above a few hundred keV.}. The column depth is fixed at 15\arcsec, estimated using the microwave source size at $\sim$5 GHz. The parameters used for fitting are the power-law spectral index of the differential electron density distribution $\delta'$, total non-thermal electron density $n_{\rm nth}$ above the low-energy cutoff, thermal plasma density $n_{\rm th}$, magnetic field strength $B$, and the angle between the line of sight and the magnetic field vector $\theta$. The fitting procedure is performed using the Markov Chain Monte Carlo (MCMC) method in order to find the global minima in the multi-parameter space (following \citealt{Chen2020}).

The best-fit source parameters and their uncertainties are presented in Figures~\ref{MW_spectra_results}(a)–(c). The magnetic field strength $B$ is at its maximum near the northern footpoint (0--5\arcsec from the starting point), reaching $\sim$500 G, and gradually decreases toward the southwest. The non-thermal electron density $n_{\rm nth}$ reaches its maxima near both the northern footpoint and the above-the-looptop region (at a distance of 12\arcsec from the northern footpoint) with a value of $\gtrsim\,3.0\times 10^{6}\,\mathrm{cm}^{-3}$. It then gradually decreases to $\sim\,3.2 \times 10^{4}\,\mathrm{cm}^{-3}$ toward the southwest, or $<$1\% of its maximum value.

%Next, we investigate the HXR nonthermal emission, which is enhanced, with its peak observed in the 50–84 keV energy channel, as shown in Figure~\ref{case_study_overview}(c). 
Next, we conduct HXR spectral analysis using a composite model consisting of an isothermal component and a nonthermal thick-target component with a power-law electron distribution using the \texttt{OSPEX} package included in the \texttt{SSWIDL} software suite \citep{Schwartz2002}. The observed X-ray spectrum and fitting results are shown in Figure~\ref{hxr_spectral_fitting}. The best-fit power-law spectral index of the differential electron flux distribution $\delta$ is $4.06\pm 0.08$ and the low-energy cutoff $E_c$ is $20.4\pm1.99$ keV. The total electron rate $\dot N$ is $1.15\pm0.27 \times 10^{35}\,\mathrm{s}^{-1}$. 

The total number of accelerated electrons produced during the jet event can be estimated using both the microwave and HXR spectral analysis results. %To minimize transport effect when comparing to the \textit{in situ} measurements, we choose to compare the number of electrons with energy above 120 keV (which is above the spectral break in the \textit{in situ} data; see next subsection). 
%For HXRs, the observations only have sufficient signal-to-noise at $\lesssim$70 keV. However, if we assume the same power-law distribution of the HXR-emitting nonthermal electrons extends to higher energies. 
For HXR-constrained electron numbers, we use three selected energy channels to calculate the total nonthermal electron rates above $E_s = 30$ keV, 50 keV, and 120 keV, which can be derived using the following equation: $\dot N_{>E_s}$ = $\dot N_{>E_c}(\frac{E_c}{E_s})^{\delta-1}$, where $E_c = 20.40$ keV is the low-energy cutoff returned from the HXR spectral analysis (Figure~\ref{hxr_spectral_fitting}). The total number of HXR-derived nonthermal electrons in the three selected energy channels is calculated using the relation $N_{>E_s}^{\text{HXR}}=\dot N_{>E_s}^{\text{HXR}}\tau_{\text{HXR}}$, where the duration $\tau_{\text{HXR}}$ is taken to be 19 seconds, based on the FWHM duration of the 25--50 keV light curve. The results are shown in the first row of Table~\ref{para_tab}. %The values of $N_{>30\,\mathrm{keV}}^{\text{HXR}}$, $N_{>50\,\mathrm{keV}}^{\text{HXR}}$, and  $N_{>120\,\mathrm{keV}}^{\text{HXR}}$, during the event are $6.7\times10^{35}$, $1.4\times10^{35}$, and $9.65\times10^{33}$, respectively.

%Here, $\tau_{\text{HXR}}$ is the duration of the presence of the HXR-emitting nonthermal electrons, taken to be 19 seconds, as derived from the FWHM of the 25–50 keV light curve. %represents the HXR emission duration. 

For the microwave estimation, we derive the total number of microwave-emitting electrons above the selected energy channel $E_s$ using the following formula:  

\begin{equation}
N_{>E_s}^{\rm MW} = \tau_{\rm MW} \cdot A \cdot \int_{E_s}^{E_{max}}{v_e\cdot{f_e}} dE
\label{Total_cal_mw}
\end{equation}
Here, $\tau_{MW}$ is the duration of the microwave emission, which is 21 seconds based on the FWHM of the 6 GHz light curve. $A$ is the cross-section area of the microwave source, set to be $15'' \times 15''$, $v_e$ is the electron velocity, and $f_e$ is the differential electron density distribution following a power-law form ($f_e=dn_e/dE\propto E^{-\delta^{'}}$). With the best-fit parameters derived from the microwave spectra, similarly, we obtain the total number of microwave-emitting nonthermal electrons above 30 keV, 50 keV, and 120 keV. %, $N_{>30\,\mathrm{keV}}^{\rm MW}$, $N_{>50\,\mathrm{keV}}^{\rm MW}$, and $N_{>120\,\mathrm{keV}}^{\rm MW}$, during the event as approximately $\sim 3.3 \times 10^{35}$, $\sim 8.8 \times 10^{34}$, and $\sim 9.6 \times 10^{33}$, respectively. 
The results are shown in the second row of Table~\ref{para_tab}. The total number of microwave-emitting electrons is consistent with that of the HXR-emitting electrons, which are of the same order of magnitude.

\begin{table*}[!ht]
    \center
    \begin{tabular}{lllll}
    \toprule
    ~ & $>$30 keV [$\times 10^{32}$] & $>$50 keV [$\times 10^{32}$] & $>$120 keV [$\times 10^{32}$] \\ 
    \midrule
    $N^{\rm HXR}$: & 6713.2 $\pm$ 2557.9 & 1406.2 $\pm$ 543.5 & 96.5 $\pm$ 39.1  \\  
    $N^{\rm MW}$: & 3323.2 $\pm$ 3650.6 & 879.7 $\pm$ 1152.8 &  95.6 $\pm$ 141.3 \\ 
    $N^{\rm SEE}$: & 5.4 $\pm 3.0$  & 3.2 $\pm$ 1.7 &  1.0 $\pm$ 0.3 \\
    $N^{\rm SEE}/N^{\rm HXR}$: & 0.08\% & 0.23\% & 0.97\%  \\
    $N^{\rm SEE}/N^{\rm MW}$: & 0.16\% & 0.36\% & 0.96\% \\

    \bottomrule
    \end{tabular}
    \caption{Total number of energetic electrons above three selected energy thresholds, 30 keV, 50 keV, and 120 keV, as derived from HXR, microwave (MW), and \textit{in situ} SEE observations.}
    \label{para_tab}
\end{table*}

\subsection{Multi-Spacecraft \textit{in situ} SEE Measurements}
On 2022 November 12, in addition to SolO/EPD, both WIND/3DP and STEREO-A/SEPT detected a significant enhancement in the \textit{in situ} energetic electron flux (Figure~\ref{in-situ_obs}). At this time of interest, the magnetic longitudinal separations between the active region and Solo and STEREO-A were both less than $20^{\circ}$. As discussed in section \ref{SEE_events_overview}, to enable a accurate comparison of peak flux among different spacecraft, here we adopt the sector-averaged flux across different viewing directions for the electron spectral analysis. In Figure~\ref{Fig_instu_electron}(a), we show the electron spectra measured by the three spacecraft derived from the peak electron flux during this event. The electron flux spectra are fitted using a broken power-law distribution. 
For the WIND observations, only five energy channels show a clear signal. Therefore, the break energy was fixed in the range of 100--120 keV according to the fitting results from STEREO-A and SolO observations. The spectral fitting results from the three spacecraft consistently indicate that the low-energy spectral indices are approximately 2, while the high-energy spectral indices are around 4, with a break energy of 100--120 keV.

%The spectral indices above the break energy ($\sim$100 keV) derived from STEREO and SolO measurements are in good agreement, with a value of $\delta_{\rm h}\approx 4.0$--4.1. The spectra below the spectral break measured by all three spacecraft are significantly flatter, with the spectral indices $\delta_{l}\lesssim 2.0$. %Considering that the electron spectral index above the break energy is close to those derived from HXR and microwave observations, 
%The flattening of the low-energy spectral index may be attributed to transport effects, such as Coulomb collision losses \citep{Lin1985} and/or plasma turbulence \citep{Kontar2009}. 

%Here, we adopt the same three selected energy channels as those used in the HXR and microwave analyses to estimate the total number of electrons. 
In addition, the three spacecraft located at widely spread longitudinal angles also allow us to estimate the angular distribution of the SEE event. Here, we fit the peak flux at 30, 50, and 120 keV, using the data observed by SolO, WIND, and STEREO-A, with a Gaussian function (following, e.g., \citealt{Richardson2014, Strauss2015, Xie2019}). In addition to the different longitudinal separations, the peak flux is strongly affected by the spacecraft's radial distance from its source $d$. In our case, SolO is located at 0.61 AU from the Sun, whereas STEREO and WIND are approximately 1 AU away. To normalize the electron flux obtained by SolO to a common heliocentric distance of 1 AU, we first scale the observed electron flux by a factor of $d^{-2}$, assuming free ballistic propagation. The scaled peak fluxes at 120 keV are shown in Figure~\ref{Fig_instu_electron}(b) as black symbols. The solid black curve represents the fitted Gaussian function. The derived FWHM of the SEE angular distribution using the multi-spacecraft measurements is 28$^{\circ}$, which is broadly consistent with the maximum separation angle of $\sim$57$^{\circ}$ for WIND to detect a SEE event (see Section~\ref{SEE_events_overview}), as well as previous statistical results \citep{Lin1974}. Some studies show that a $d^{-3}$ dependence may be more likely due to transport effects \citep{Rodriguez-Garcia2023}. The gray dashed line in Figure~\ref{Fig_instu_electron} (b) represents the fitted Gaussian function using the peak flux scaled by the $d^{-3}$ relation instead. In this case, the FWHM of the best-fit Gaussian function is 40 degrees. Owing to the similar spectral shapes observed by the three spacecraft, similar results for the angular extent can be derived if we choose other energy channels for this analysis. 

The \textit{in situ} SEE spectra and estimated angular extent are, in turn, used to estimate the total number of electrons escaped into the interplanetary space (see, e.g., \citealt{James2017}). By integrating the event duration of $\sim$2.5 hours, defined as the period during which the flux at least exceeds three sigma above the background, the total number of energetic electrons above 30 keV, 50 keV, and 120 keV that manage to escape into the interplanetary space are shown in the third row of Table~\ref{para_tab}. %about $5.4\times 10^{32}$,  $3.2\times 10^{32}$, and $1.0\times 10^{32}$, respectively. 
Compared to the estimated total number of HXR- and microwave-emitting electrons near the solar surface, the ratio of escaped electrons to those near the solar surface is between 0.1 and 1\%. These values are extremely small and are consistent with previous findings.

\section{Discussion and conclusion} \label{discussion}

\begin{figure*}
    \includegraphics[width=1\textwidth]{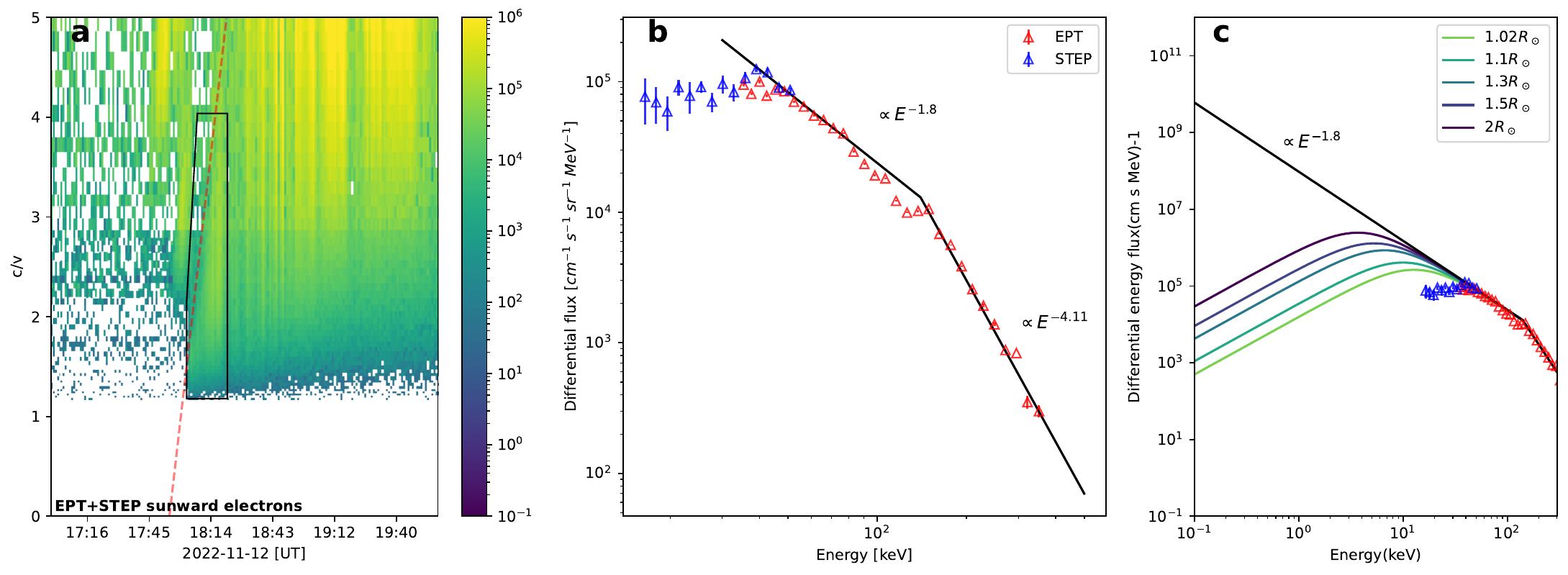}
    \caption{Combined energetic electron spectrum of the 2022 November 12 SEE event, as observed by the sunward detectors of the EPT and STEP instruments onboard SolO. (a) Combined SEE spectrogram with background subtracted. The vertical axis represents the energetic electron energy converted to $c/v$, where $c$ is the speed of light and $v$ is the electron velocity. %The red dashed line indicates the estimated onset time, determined using the HXR/microwave emission peak time and assuming the electrons travel ballistically along the Parker spiral line. 
    The black box marks the SEE event of interest, with a clear velocity dispersion feature. %The red dashed line indicates the best-fit velocity-dependent onset time, determined using the HXR/microwave emission peak time and assuming the electrons travel ballistically along the Parker spiral line. 
    (b) Combined EPT and STEP background-substracted electron spectra with $3\sigma$ background error bars derived from the energy-dependent peak times within the black box in panel (a). The blue and red points represent measurements made by STEP and EPT, respectively. (c): Coulomb collision model fits of the observed SEE spectrum. The curves in different colors represent the model electron spectra with different injection heights. %The red and blue points show the observed electron spectra, the same as the spectra in panel (b).
    }
    \label{ept_step_energy_loss_model}
\end{figure*}

In this paper, we started from the survey presented by \citet{Lario2024} on a series of SEE events occurring from 2022 November 9 to 2022 November 14, incorporating both \textit{in situ} measurements and multi-wavelength remote-sensing observations. We placed special focus on 25 SEE events originating from AR 13141, considering its favorable magnetic connectivity to the SolO spacecraft. We found that:% based on the summary table reported by \citet{Lario2024}, analyzing them from both temporal and spatial perspectives. Here's the summary of our main findings about event survey.

\begin{itemize}
    \item Nearly all 25 SEEs are associated with interplanetary type III radio bursts with only one event showing a time delay.
    \item 24 out of the 25 SEEs are related to solar jets or eruption processes.
    %\item If there are neither type III radio bursts nor jet or eruption events, then no SEEs are detected.
    \item 17 out of the 25 SEEs with available X-ray imaging data from SolO/STIX are concentrated in a distinct recurrent jet region located at the southwest edge of AR 13141 that features mixed polarities. %These correlations strongly suggest that the origin of SEEs during November 10–12, 2022, is highly likely to be associated with solar jets.
    \item The recurrent jet source region has a very small ($<20^{\circ}$) longitudinal separation between the ballistically projected footpoints of SolO. These footpoints are located close to a narrow region with direct access to open field lines, favorable for particles escape to the interplanetary space.  
\end{itemize}

The above findings strongly suggest that jet events in this specific region, which are accompanied by interplanetary type III radio bursts, share a common origin with the corresponding in-situ SEE event. This connection enables a direct comparison of the energetic electron population near the solar surface with the population that enters the interplanetary space. Next, we focus on a specific jet-associated SEE event that occurred on 2022 November 12 at around 18 UT, which has excellent coverage by a suite of remote-sensing and \textit{in situ} instruments including SolO, EOVSA, SDO, WIND, and STEREO-A. In particular, joint HXR and microwave imaging spectroscopy offered by SolO/STIX and EOVSA provides comprehensive diagnostics of the nonthermal electron distribution over a wide energy range. Meanwhile, \textit{in situ} measurements made by multiple spacecraft (SolO, WIND, and STEREO-A) located at various heliographic longitudes and distances offer a great opportunity to constrain the angular distribution of energetic electrons entering the interplanetary space, derived to have a FWHM angle of 28$^{\circ}$ assuming free propagation.

The \textit{in situ} SEE spectra observed by three different spacecraft at different heliocentric longitudes all exhibit a broken power-law distribution with similar power-law indices and break energies around 100--120 keV. Only the absolute flux varies with each spacecraft's angular separation from the source. This spectral uniformity across longitudes suggests that the broken power-law distribution may be an intrinsic property of the accelerated SEE population entering interplanetary space. %suggesting that the interplanetary transport effects did not significantly influence electrons with energies above 30 keV during this event. 

For the HXR spectral analysis ($\sim$20--100 keV), which serves as a remote diagnostic of energetic electrons near the solar surface, we used a single power-law distribution. A broken power-law was not employed because the data show no sign of a break in this range.
For microwave spectral analysis, we also adopted a single power-law form for the nonthermal electron distribution mainly because the microwave data alone do not have sufficient constraints for determining a more complicated distribution with additional free parameters (such as a broken power-law form; see, however, \citealt{Chen2021} for a successful example when both microwave and HXR data could be utilized to jointly fit the same nonthermal coronal source). %in order to reduce the number of free parameters and to account for the limited high-energy range ($>50$ keV) of the hard X-ray data. 
Nevertheless, the spatially resolved microwave spectral fitting results suggest that the spectral index of the microwave-emitting electrons near the footpoint (the first point in Figure~\ref{MW_spectra_results}(c)) is $\delta \approx \delta'-0.5 \approx 4.0$, consistent with the HXR fitting results. A harder spectrum of $\delta\approx3.1$ is observed near the looptop (at a distance of $\sim\!15''$ in Figure~\ref{MW_spectra_results}(c)).
A harder electron spectrum at the looptop has also been reported in the literature \citep[e.g.,][]{Costa2005,Mondal2024}, which may be attributed to more effective trapping of high-energy electrons in the coronal acceleration region due to turbulent pitch-angle scattering \citep{Kontar2014,Musset2018}.
%\textbf{The softer electron spectrum at the footpoint is possibly due to the propagation effects experienced by the accelerated electrons as they travel from the loop-top region to the footpoint. Higher-energy electrons are more efficiently trapped than lower-energy electrons} \citep{Kuznetsov2015, Musset2018, Mondal2024}. 
Interestingly, the energetic electron spectrum in the region further away from the looptop region close to the base of the EUV jet spire, despite having a much smaller number density, exhibits an even harder spectrum with $\delta\approx 2.5$, approaching the spectral index derived from \textit{in situ} observations in the 30--100 keV range.

Next, in Section~\ref{case_study_overview_sec}, we have estimated the total number of HXR/microwave-emitting nonthermal electrons near the solar surface and compare it to those entering the interplanetary space measured by multi-spacecraft \textit{in situ} measurements. %There, we determine the ratio of the electrons entering the interplanetary space to those near the solar surface to be 0.1--1\%. %Such an extremely small ratio is consistent with the values reported earlier \citep{Krucker2007, Wang2021, Dresing2021}. As we discussed in Section~\ref{sec:intro}, one possible reason is that those entering interplanetary space could be strongly affected by transport effects, which may significantly reduce the \textit{in situ} electron flux.
%could influence the electron's peak value, as discussed, which would also affect the ratio. 
%Indeed, our observed \textit{in situ} electron distribution with a broken power-law form suggest that transport effects are likely present to alter the electron spectra below the break energy at $\sim$100 keV (e.g., \citealt{Kontar2009, Droge2016, Strauss2020}). To mitigate such effects, when making the comparison with the HXR/microwave-derived electron number, we select the energy channels above 120 keV, which are much less affected by transport effects as evidence by the highly consistent spectral indices measured by SolO, WIND, and STEREO-A across a wide range of heliographic longitudes. 
We notice that the estimation of the total number of escaping electrons based on \textit{in situ} measurements can be strongly affected by the angular extent of nonthermal electrons entering interplanetary space. For single spacecraft measurements, this angular extent is essentially unknown and has to be assumed. A nominal angular width of 30$^{\circ}$ is often assumed (e.g., \citealt{Krucker2007}) based on statistical studies of the longitudinal distribution of SEE-associated flare events observed from Earth \citep{Lin1974,Reames1999}. However, if the angular width were larger than the assumed value by a factor of 3, the total number of electrons escaping to interplanetary space can increase by nearly an order of magnitude. Indeed, in one jet event, using microwave imaging spectroscopy observations of a type III radio burst in $\sim$1--3 GHz, \citet{wang2023} reported that the angular extent of the type-III-burst-emitting electrons escaping along open field lines is much broader than the EUV jet spire, reaching $\sim$90$^{\circ}$. %The total number of electrons can vary by a factor of $\sim$10 due to uncertainties in the width of the angular distribution, typically assumed to range from $30^{\circ}$ to $90^{\circ}$. 
In this study, by utilizing observations from three spacecraft located at various longitudinal positions, we have a more robust constraint of the angular extent of the escaping electrons to be from 28--40$^{\circ}$, thus greatly reducing the associated uncertainties of the derived ratio of escaped vs. retained energetic electrons. 

Based on the improved constraints discussed above, we conclude that the fraction of energetic electrons that escape into the interplanetary space is extremely small, only 0.1--1\% of the microwave/HXR-emitting electrons near the solar surface, consistent with previous reports.
%Previous studies (e.g., \citealt{Krucker2007, Wang2021}) have conducted statistical analyses to investigate the connection between HXR emissions and \textit{in situ} measurements. 
As introduced earlier, one interpretation posits that the \textit{in situ} SEEs originate from an initial acceleration site high in the corona; thereby, they are much lower in number and absence of strong Coulomb collision losses \citep{Wang2021}. Our observations do not favor this scenario. %Here, we examine the lower-energy electrons detected by another sensor, the SupraThermal Electrons and Protons (STEP) instrument onboard EPD, which measures electrons with energies extending down to 4 keV. 
Figure~\ref{ept_step_energy_loss_model}(a) shows SolO/EPT observations of the SEE spectrogram over a wide energy range from $\gtrsim$10 keV to 400 keV. Unlike the electron spectral analysis across different spacecraft, to ensure consistency when combining the electron flux measurements from the SupraThermal Electrons and Protons (STEP) instrument with those from SolO/EPD, we adopt the data from sunward sensor viewing directions. A clear velocity-dispersion is seen (red dashed line), which is due to higher-energy (or faster) electrons arriving at the spacecraft at an earlier time. Assuming that the electrons propagate along the nominal Parker spiral to SolO, we find that the SEEs are released near the peak time of the microwave and HXR emissions, confirming the common origin of the \textit{in situ} SEEs and the HXR/microwave-emitting energetic electrons. %The onset times of the energy channels inside the black box align well with the estimated onset time (red dashed line), displaying clear velocity dispersion. 
From there, we extract peak electron fluxes at least three sigma above the background level to represent the \textit{in situ} electron spectra shown in panel (b). We find that the electrons at low energies (10--40 keV) display a flatter spectrum, which differs from the cases reported by \citet{Wang2021}, which showed the same power-law extends to extremely low energies ($<$5 keV). The observed spectrum with a flat or positive spectral slope suggests that these low-energy electrons originated from the very low corona and have experienced substantial energy loss during their propagation to the spacecraft.

Following earlier work \citep{Lin1985,Wang2021,wang2023}, we model the observed SEE spectrum by injecting energetic electrons from various coronal heights with a power-law distribution with a spectral index of $\delta = 2.0$ up to the observed break energy of 120 keV. They lose energy during their propagation to the spacecraft as a result of Coulomb collisions in an energy-dependent manner. Figure~\ref{ept_step_energy_loss_model}(c) shows the modeled electron spectra with different injection heights (1.02~$R_{\odot}$, 1.1~$R_{\odot}$, 1.3~$R_{\odot}$, 1.5~$R_{\odot}$, and 2~$R_{\odot}$) using the same density model as in \citet{wang2023} but with the scaling factor adjusted to five times the Newkirk model \citep{Newkirk1967} in the low corona. As expected, the low-energy electron spectrum's turn-over energy shifts to higher energies for lower injection heights, as the electrons undergo more Coulomb collision losses. %\textbf{We note that, with an injection height of 1.02 $R_{\odot}$, the observed turnover energy is still higher than that modeled assuming free propagation. This implies that the density may be higher than the adopted model or electrons experience additional transport effects that increase their effective propagation distance (e.g. \citealt{Kontar2009, Strauss2017})}. 
Unlike the cases reported in \citet{Wang2021}, we find that the observed electron spectrum with a turnover energy of $\sim\!40$ keV is best described by the model with a very low injection height of $<$1.02~$R_{\odot}$, or $<20''$ above the solar surface, consistent with implications from our remote-sensing results (c.f., illustration in Figure~\ref{fig:cartoon}). We note that, even with a low injection height of 1.02~$R_{\odot}$, the observed turnover energy is still higher than that modeled assuming free propagation. This implies that the low-coronal density may be greater than the adopted model or electrons experience additional transport effects that increase their effective propagation distance (e.g. \citealt{Kontar2009, Strauss2017}).  %Assuming the electrons are affected by Coulomb collision losses, the observed electron spectra is close to the model assuming a source height of 1.02 solar radii, corresponding to a region with a density approximately five times that of the Newkirk density model in the low corona. 
%Therefore, in our case, there is no clear evidence supporting secondary acceleration of downward-propagating electrons. 
Therefore, we conclude that the \textit{in situ} SEEs in our event likely share a common origin with the HXR- and microwave-emitting electrons near the solar surface, but are not involved in the secondary acceleration process. %as supported by the similar electron spectral indices derived from both remote-sensing and \textit{in situ} (electrons $>$ 120 keV) observations.  Transport effects could play a key role in the observed changes in the electron spectra at the low energies below the break energy.

Another possibility interpretation favored by our observations is that the population of upward-propagating electrons is significantly depleted before reaching interplanetary space because of effective trapping in the electron acceleration region presumably located above the mini-arcade at the jet base. A promising trapping mechanism may be strong diffusion due to turbulence-induced scattering. For example, previous modeling studies have shown that, with strong diffusion, the nonthermal electron density can be reduced by two orders of magnitude at a distance of 10-20 Mm away from the acceleration region \citep{Musset2018,Chen2024}. This depletion results in an extremely small ratio of the escaping electrons compared to the HXR/microwave-emitting electrons. This possibility is strongly supported by our spatially resolved microwave spectra analysis results, which show that the nonthermal electron density is only about two orders of magnitude at the southwestern-most point where they enter the open field lines (see Figure~\ref{MW_spectra_results}(a)) compared to its peak value near the mini-postflare arcade. %If we estimate the total number of electrons above 120 keV using Equation~\ref{Total_cal_mw} and the density at the southwesternmost point of $3.4 \times 10^{4} cm^{-3}$, the total number of electrons drops to $\sim1.9 \%$ compared to the MW/HXR-emitting electrons derived from the peak density before their release into the interplanetary medium. 
Coincidentally, this drop in density of the nonthermal electrons entering open field lines matches the interplanetary-space-to-solar-surface electron ratio by the same order of magnitude ($\sim$1\%). % This total number of electrons is nearly the same as the measured escaping electrons that reach the spacecraft.

Our observations of the small eruptive jet event are consistent with the scenario proposed by \citet{Chen2024} for the gradual phase of a major eruptive solar flare (X8.2 class). In our case,  the majority of nonthermal electrons may be accelerated and trapped in the above-the-loop-top region above the mini-postflare arcade, where a high concentration of microwave-emitting nonthermal electrons coincides with a reduced magnetic field strength, referred to as a ``magnetic bottle'' \citep{Chen2020}. %In our case, the blowout jet eruption consists of two stages. The initial internal reconnection is similar to the standard CSHKP flare scenario \citep{Carmichael1964, Sturrock1966}, and a similar concentration of nonthermal electrons in the above-the-loop-top magnetic bottle region is observed. 
%Despite the limited spatial coverage of the spectral fitting, 
Our results indicate that the region located 10–15\arcsec\ from the northern point exhibits these characteristics, including a lower magnetic field strength and a peak in nonthermal electron density. As the distance increases toward the southwest, the nonthermal electron density continuously decreases due to the effective trapping in the magnetic bottle region, possibly by turbulence and/or strong magnetic mirroring, resulting in a very small number density of nonthermal electrons (two orders of magnitude lower than the peak density). The location of these microwave-emitting electrons coincides with the base of the jet spire above the reconnection site, which may find a path to enter the interplanetary space via the open field lines, giving rise to a small population of \textit{in situ} SEEs.%\textbf{However, while this interpretation represents a primary explanation for the small fraction of escaping electrons, additional contributions from other factors discussed earlier, such as transport effects and secondary acceleration processes, cannot be fully excluded.}

Our study benefits from the selection of a favorable time period, during which a series of SEEs shows a close association with type III radio bursts and solar jets originating from an active region with excellent magnetic connectivity. This association provides us with strong confidence in the origin of these SEEs, allowing us to further investigate their acceleration and transport processes. The joint use of HXR and microwave observations probes a wide range of electron energies, allowing us to minimize the transport effects when comparing with \textit{in situ} measurements. Additionally, multi-spacecraft \textit{in situ} measurements of SEEs provide a tight constraint on the angular distribution of the energetic electrons, giving us a more accurate estimate for the electrons entering the interplanetary space. More importantly, the simultaneous microwave imaging spectroscopy capability offered by EOVSA, when complemented by EUV, HXR, and magnetic field data, enables us to spatially resolve the nonthermal electron distribution in the core jet region, allowing us to shed new light on the small escaping electron population problem. % and offers new insights into the spatial distribution of electrons at the acceleration site. 
In the future, next-generation telescopes, such as the Frequency Agile Solar Radio telescope (FASR; \citealt{Gary2023, Chen2023}), with significantly higher dynamic range, sensitivity, and angular resolution, are expected to greatly enhance our understanding of the acceleration and transport of SEE events.

\section*{Acknowledgments}
    The authors acknowledge the Solar Orbiter Energetic Particle Detector (EPD) team for their guidance in analyzing the electron data. M.~W., B.~C., and S.~Y. were supported by NSF SHINE grant AGS-2229338 and NASA grants 80NSSC24K1242 and 80NSSC20K1282 to New Jersey Institute of Technology. J.~L. acknowledges NSF grant, AGS-2114201, and NASA grant, 80NSSC24K0258.  H.~W. acknowledges NSF Grant AGS-2209064, NASA grants 80NSSC20K1282 and 80NSSC24M0174.  M.~Wickline acknowledges support by NSF REU program under grant AGS-2050792 ``REU Site: Solar, Terrestrial, and Space Weather Sciences at New Jersey Institute of Technology.'' The Expanded Owens Valley Solar Array (EOVSA) was designed, built, and is now operated by the New Jersey Institute of Technology as a community facility. The EOVSA operations are supported by NSF grants AGS-2130832, AGS-2436999, and NASA grant 80NSSC20K0026 to NJIT.

%\appendix

%\bibliography{sample631}{}
%\bibliographystyle{aasjournalv7}

\end{document}